\documentclass{article}

\usepackage{arxiv}

\usepackage[utf8]{inputenc} 
\usepackage[T1]{fontenc}    
\usepackage{lmodern}        
\usepackage{hyperref}       
\usepackage{url}            
\usepackage{booktabs}       
\usepackage{amsfonts}       
\usepackage{nicefrac}       
\usepackage{microtype}      
\usepackage{graphicx}

\title{Spatial Joint Models through Bayesian Structured Piece-wise
Additive Joint Modelling for Longitudinal and Time-to-Event Data}

\author{
    Anja Rappl
    \thanks{corresponding author}
   \\
    Institute of Medical Informatics, Biometry and Epidemiology \\
    Friedrich-Alexander Universität Erlangen-Nürnberg \\
  Germany \\
  \texttt{\href{mailto:anja.rappl@fau.de}{\nolinkurl{anja.rappl@fau.de}}} \\
   \And
    Thomas Kneib
   \\
    Chair of Statistics \\
    Georg-August-Universität Göttingen \\
  Germany \\
  \texttt{} \\
   \And
    Stefan Lang
   \\
    Department of Statistics \\
    Universität Innsbruck \\
  Austria \\
  \texttt{} \\
   \And
    Elisabeth Bergherr
   \\
    Chair of Spatial Data Science and Statistical Learning \\
    Georg-August-Universität Göttingen \\
  Germany \\
  \texttt{} \\
  }

\providecommand{\tightlist}{%
  \setlength{\itemsep}{0pt}\setlength{\parskip}{0pt}}

\newlength{\cslhangindent}
\newlength{\csllabelwidth}
\newlength{\cslentryspacingunit} 
  {}%
  {\par}
\newenvironment{CSLReferences}[2] 
 {
  \setlength{\parindent}{0pt}
  \ifodd #1
  \let\oldpar\par
  \def\par{\hangindent=\cslhangindent\oldpar}
  \fi
  \setlength{\parskip}{#2\cslentryspacingunit}
 }%
 {}
\usepackage{calc}

\usepackage{array}
\usepackage{booktabs}
\usepackage{cancel}
\usepackage{colortbl}
\usepackage{float}
\usepackage{longtable}
\usepackage{makecell}
\usepackage{multirow}
\usepackage{pdflscape}
\usepackage{siunitx}
\usepackage{tabu}
\usepackage{threeparttable}
\usepackage{threeparttablex}
\usepackage[normalem]{ulem}
\usepackage{wrapfig}
\usepackage{xcolor}
\usepackage{arxiv}
\usepackage{amsmath}
\usepackage{booktabs}
\usepackage{longtable}
\usepackage{array}
\usepackage{multirow}
\usepackage{wrapfig}
\usepackage{float}
\usepackage{colortbl}
\usepackage{pdflscape}
\usepackage{tabu}
\usepackage{threeparttable}
\usepackage{threeparttablex}
\usepackage[normalem]{ulem}
\usepackage{makecell}
\usepackage{xcolor}
\begin{document}
\maketitle

\begin{abstract}
Joint models for longitudinal and time-to-event data have seen many
developments in recent years. Though spatial joint models are still rare
and the traditional proportional hazards formulation of the
time-to-event part of the model is accompanied by computational
challenges. We propose a joint model with a piece-wise exponential
formulation of the hazard using the counting process representation of a
hazard and structured additive predictors able to estimate (non-)linear,
spatial and random effects. Its capabilities are assessed in a
simulation study comparing our approach to an established one and
highlighted by an example on physical functioning after cardiovascular
events from the German Ageing Survey. The Structured Piece-wise Additive
Joint Model yielded good estimation performance, also and especially in
spatial effects, while being double as fast as the chosen benchmark
approach and performing stable in imbalanced data setting with few
events.
\end{abstract}


\hypertarget{introduction}{%
\section{Introduction}\label{introduction}}

Biometrical studies often capture time-to-event and longitudinal data on
the same topic simultaneously. Frequently used examples are the count of
CD4 lymphocytes in HIV-positive patients and their time till onset of
AIDS (Faucett and Thomas 1996; Wulfsohn and Tsiatis 1997; Rizopoulos
2011) or the level of serum bilirubin and other liver biomarkers in
primary biliary cirrhosis patients and time to death (Crowther, Abrams,
and Lambert 2013; Hickey et al. 2018). Other examples include PSA cancer
marker and progression to recurrence of prostate cancer (Jacqmin-Gadda
et al. 2010), autoantibody titers in children preceding the onset of
Type 1 diabetes (Köhler, Beyerlein, et al. 2017) or physical functioning
after a cardiovascular event and death (Rappl, Mayr, and Waldmann 2022).
Separate analysis of these longitudinal and time-to-event outcomes leads
to biased estimates and to avoid this both should be modelled jointly.
These joint models consist of two submodels: A longitudinal submodel and
a survival submodel with both being linked through an association
parameter.\\
While Wulfsohn and Tsiatis (1997) and Henderson, Diggle, and Dobson
(2000) proposed to maximize the likelihood of a joint model via an
Expectation-Maximization (EM) algorithm, Faucett and Thomas (1996) used
a Bayesian Gibbs-sampling approach. In recent years advances have been
made into statistical boosting (Waldmann et al. 2017; Griesbach, Groll,
and Bergherr 2021). Software is available for all three estimation
approaches across various statistical computation platforms, of which
especially \texttt{R} hosts a number of well-established packages such
as \texttt{JM} (Rizopoulos 2010), \texttt{JMbayes} (Rizopoulos 2016),
\texttt{joineRML} (Hickey et al. 2018) and \texttt{bamlss} (Umlauf et
al. 2021). Comparisons of selections of available software can be found
in Yuen and Mackinnon (2016) and Rappl, Mayr, and Waldmann (2022).\\
Traditionally the longitudinal submodel of a joint model is a linear
mixed model (LMM) and the time-to-event submodel is a proportional
hazards (PH) model, though other variants are possible. Depending on the
scaling of the longitudinal outcome a generalized linear mixed model
(GLMM) (Faucett, Schenker, and Elashoff 1998; Rizopoulos et al. 2008;
Viviani, Alfó, and Rizopoulos 2014) or quantile regression model (Y.
Huang and Chen 2016; Zhang et al. 2019) might be better suited. An
alternative to the PH-models in the time-to-event submodel are
accelerated failure time models (Tseng, Hsieh, and Wang 2005; Y. Huang
and Chen 2016) and in certain data situations competing risks models are
best suited (X. Huang et al. 2011; Andrinopoulou et al. 2014; Blanche et
al. 2015). Also models with multivariate longitudinal outcomes are in
use (Lin, McCulloch, and Mayne 2002; Rizopoulos and Ghosh 2011; Mauff et
al. 2020) as are location-scale models (Barrett et al. 2019). Köhler,
Umlauf, et al. (2017) expanded joint models to structured additive joint
models with possibly smooth random effects and established non-linear
association structures (2018) both via a Bayesian flexible
tensor-product approach using Newton-Raphson procedures and
derivative-based Metropolis-Hastings sampling. A good historic overview
on joint models can be found in Tsiatis and Davidian (2004), while
Alsefri et al. (2020) give a concise summary of recent developments in
Bayesian joint models in particular.\\
Still joint models with a spatial component are rare. Martins, Silva,
and Andreozzi (2016) and Martins, Silva, and Andreozzi (2017) have
described estimation of a Bayesian joint model with a spatial effect and
a Weibull baseline hazard using OpenBUGS and WinBUGS respectively. The
above mentioned Bayesian tensor-product approach by Köhler, Umlauf, et
al. (2017) implemented in the \texttt{R} package \texttt{bamlss} also
has the capability of estimating spatial joint models. In terms of model
formulation both methods have in common that they use a PH-model for the
survival submodel. However, assuming a parametric baseline hazard such
as a Weibull hazard can be restrictive and derivative-based
Metropolis-Hastings algorithms are computationally expensive as well as
may prove sensitive towards data with few events.\\
Therefore, in this paper we propose a Bayesian joint model with a
structured additive LMM for the longitudinal outcome, but exchange the
time-to-event submodel for a piece-wise additive mixed model (PAMM). The
latter has been suggested by Bender, Groll, and Scheipl (2018) for
modelling survival times based on the proportionality of a time-to-event
process with a Poisson-distributed count process (Friedman 1982) thus
expanding the available options for time-to-event models
(e.g.~accelerated failure times, competing risk). This formulation
allows for estimation of the baseline hazard without any assumptions
about its distributional form and is similarly flexible to the Köhler,
Umlauf, et al. (2017) model with respect to the inclusion of
(non-)linear, spatial and random effects. At the same time, it reduces
runtimes by about 50\% (compared to an established method) and has
proven stable in imbalanced data settings with few events.\\
The rest of the paper is structured as follows: In the next section, the
methodology of piece-wise additive joint models is described in more
detail and our extension of the concept is explained. In section three
the results of a simulation study comparing our approach to an
established one to proof the feasibility of the model formulation, its
ability to estimate spatial effects and its runtime performance. We then
apply this method to an example of physical functioning from the German
Aging Survey. Section five concludes with some final remarks and further
technical details can be found in the Appendix.

\hypertarget{methods}{%
\section{Methods}\label{methods}}

\hypertarget{theoretical-background}{%
\subsection{Theoretical background}\label{theoretical-background}}

In its original form the Joint Model assumes a linear mixed model (LMM)
for the longitudinal outcome and a proportional hazards model (PH) for
the time-to-event outcome (Wulfsohn and Tsiatis 1997; Faucett and Thomas
1996; Henderson, Diggle, and Dobson 2000).\\
Let \(\boldsymbol{y}\) denote the vector of longitudinal outcomes across
all individuals \(i = \{1, \dots, n\}\) and observations times points
\(t\). Further, let \(\boldsymbol{\lambda}(t)\) be the vector of
individual specific risks to experience an event at time \(t\)
proportional to the baseline hazard \(\lambda_0(t)\) and based on the
observed event or censoring times \(\boldsymbol{T}\) and event indicator
\(\boldsymbol{\delta}\). Then in its most generic variant the original
joint model takes the form \begin{align}
\boldsymbol{y}(t) & = \boldsymbol{\eta}_\text{l}(t) + \boldsymbol{\eta}_{\text{ls}}(t) + \varepsilon, \quad \varepsilon \sim \text{N}(0, \sigma^2\boldsymbol{I}) \label{rappl:JMlong} \\
\boldsymbol{\lambda}(t) & = \lambda_0(t) \exp\{\boldsymbol{\eta}_\text{s} + \alpha \boldsymbol{\eta}_{\text{ls}}(t)\} \label{rappl:JMsurv},
\end{align} where \(\boldsymbol{\eta}_\text{s}\) and
\(\boldsymbol{\eta}_\text{l}\) are survival and respectively
longitudinal submodel specific predictors and
\(\boldsymbol{\eta}_{\text{ls}}\) is the shared predictor, via which
both model parts are connected. The parameter \(\alpha\) quantifies the
association between the longitudinal and the time-to-event outcome. In
the following, the addendum \(\cdot(t)\) to denote time-varying
predictors is dropped for ease of notation, since the subsequent
concepts can be applied to time-varying and -constant predictors alike.
Also note that, while it is theoretically possible to estimate
time-varying survival predictors \(\boldsymbol{\eta}_\text{s}(t)\), the
time-varying covariates included in that predictor may be prone to
measurement error and it is therefore in most cases better to model them
jointly.\\
The predictors \(\boldsymbol{\eta}_\cdot\) are additive and may include
(non-)linear, geographical or random effects of potentially time-varying
covariates \(\boldsymbol{x}_k(t)\),
i.e.~\(\boldsymbol{\eta}_{\cdot} = \sum_{k=1}^{p_{\cdot}} f_k(\boldsymbol{x}_k(t))\),
where \(f_k\) is a function representing the respective effect and
\(p_\cdot\) denotes the predictor specific number of covariates.
Restrictions apply to random effects, which need to be part of the
shared predictor \(\boldsymbol{\eta}_{\text{ls}}\), and geographical
effects, of which there can only be one in the model for identifiability
reasons.\\
Reformulating predictor \(\boldsymbol{\eta}_\cdot\) in matrix notation
yields \begin{equation}\label{matrixNot}
    \boldsymbol{\eta}_{\cdot} = \boldsymbol{Z}_1\boldsymbol{\gamma}_1 + \cdots + \boldsymbol{Z}_{p_{\cdot}}\boldsymbol{\gamma}_{p_{\cdot}},
\end{equation} where \(\boldsymbol{Z}_k\) is an effect appropriate
design matrix and \(\boldsymbol{\gamma}_k\) a vector of corresponding
effect coefficients. For the Bayesian estimation of this model the
generic prior for the coefficients \(\boldsymbol{\gamma}_k\) is
proportional to a normal distribution with zero mean, variance
\(\sigma_{\gamma_k}^2\) and penalty matrix \(\boldsymbol{K}_k\)
\begin{equation}\label{rappl:prior}
    p\left(\boldsymbol{\gamma}_k \mid \sigma_{\gamma_k}^2\right) \propto \left(\sigma_{\gamma_k}^2\right)^{-\text{rk}(\boldsymbol{K}_k)} \exp \left\{-\frac{1}{2\sigma_{\gamma_k}^2} \boldsymbol{\gamma}_k'\boldsymbol{K}_k \boldsymbol{\gamma}_k \right\}.
\end{equation} For non-linear and spatial effects the penalty matrix
\(\boldsymbol{K}_k\) is rank deficient and as a result prior
\eqref{rappl:prior} is partially improper.\\
\textbf{Linear effects.} For a vector
\(\boldsymbol{\gamma}_k = (\gamma_{k_1}, \dots, \gamma_{k_{J_k}})'\) of
\(J_k\) linear fixed effects the penalty matrix \(\boldsymbol{K}_k\) is
an \(J_k \times J_k\) identity matrix \(\boldsymbol{I}_{J_k}\) reducing
\eqref{rappl:prior} to a \(J_k\)-variate normal distribution. An
alternative is to set
\(p(\gamma_{k_j} \mid \cdot) \propto const \quad \forall j= 1, \dots, J_k\).
The corresponding design matrix \(\boldsymbol{Z}_k\) is a matrix of
covariates of order \(n \times J_k\), where \(n\) denotes the number of
observations.\\
\textbf{Random effects.} In the case of joint models random effects
appear in the shared predictor exclusively. Thus let \(n\) be the number
of individuals and \(n_i\) be the number of observations per individual
\(i\), so that the total number of observations amounts to
\(N = \sum_{i=1}^n n_i\). Further, let \(\boldsymbol{u}_i\) be a vector
of observations (or \(\boldsymbol{1}\) for random intercepts) of length
\(n_i\) specific to individual \(i\). Then \(\boldsymbol{Z}_k\) is a
matrix of vectors \(\boldsymbol{u}_i\) of order \(N \times n\),
i.e.~\(\boldsymbol{Z}_k = \text{diag} (\boldsymbol{u}_1, \dots, \boldsymbol{u}_n)\)
and \(\boldsymbol{\gamma}_k\) is a vector of random effects \(b_i\) of
length \(n\), \(\boldsymbol{\gamma}_k = (b_1, \dots, b_n)'\). The
penalty matrix \(\boldsymbol{K}_k\) then is an \(n \times n\) identity
matrix \(\boldsymbol{I}_{n}\).\\
\textbf{Non-linear effects.} Modelling non-linear effects follows the
Bayesian P-spline approach with \(\boldsymbol{Z}_k\) being a matrix of
B-spline basis functions evaluated at observations \(x_i(t)\). Then
\(\boldsymbol{\gamma}_k\) is a vector of corresponding basis
coefficients. The common choice of prior for these basis coefficients is
a first or second order random walk. This is achieved by setting the
penalty matrix \(\boldsymbol{K}_k\) equal to
\(\boldsymbol{D}'\boldsymbol{D}\),
i.e.~\(\boldsymbol{K}_k=\boldsymbol{D}'\boldsymbol{D}\), where
\(\boldsymbol{D}\) is a matrix of first or second order differences.\\
\textbf{Spatial effects.} For spatial effects \(\boldsymbol{Z}_k\) is
assumed to be an \(n \times S\) incidence matrix (potentially also
\(N \times S\), for spatio-temporal observations) with an entry of \(1\)
if observation \(i \; \forall \; i= 1,\dots, n\) originates from
location \(s\; \forall \; s= 1,\dots, S\) with \(S\) unique locations
and \(0\) otherwise. The corresponding coefficients
\(\boldsymbol{\gamma}_k\) follow a Markov random field (MRF) prior
achieved via the penalty matrix \(\boldsymbol{K}_k\).
\(\boldsymbol{K}_k\) is an adjacency matrix of order \(S\times S\) with
entries as the number of neighbours \(|n(s)|\) only when locations \(s\)
and \(r\) are neighbours (\(s \sim r\)) of the form \begin{equation*}
\boldsymbol{K}_k[s,r] = 
       \begin{cases}
       -1 & \text{if } s\neq r, s \sim r \\
       0 & \text{if } s\neq r, s \not\sim r \\
       |n(s)| &\text{if } s=r
        \end{cases}.
\end{equation*} The variance parameters of the coefficient distributions
\(\sigma_{\gamma_k}^2\) as well as the model variance
\(\sigma_{\varepsilon}^2\) will a priori follow inverse gamma
distributions, in particular \begin{gather*}
\sigma_{\gamma_k}^2 \sim \text{IG}(a,b) \quad \text{and}\\
\sigma_{\varepsilon}^2 \sim \text{IG}(a_0,b_0).
\end{gather*}

\hypertarget{sec:PAMM}{%
\subsection{The piecewise expontential representation of the
time-to-event submodel}\label{sec:PAMM}}

The idea behind a PH-model is that an individual's hazard at time \(t\)
is determined by an individual specific deviation of an underlying
baseline hazard \(\lambda_0(t)\) at time \(t\). In mathematical notation
a generic PH-model looks similar to \eqref{rappl:JMsurv} and takes the
form \begin{equation*}
\boldsymbol{\lambda}(t) = \lambda_0(t)\exp\{\boldsymbol{\eta}\},
\end{equation*} where \(\boldsymbol{\eta}\) represents an unspecified
predictor. The aim of estimating such a model then is quantifying the
coefficients governing \(\boldsymbol{\eta}\) and determining
\(\lambda_0(t)\) over time \(t\) given the times to event
\(\boldsymbol{T}\) and the events \(\boldsymbol{\delta}\). Now this
approach can be re-written as an equivalent log-linear Poisson-model.
This is achieved by dividing the continuous observation time
\(t = (0, t_{max}]\) into \(J\) intervals and counting the events
\(\boldsymbol{\delta}_j\) in any given interval \(j\). The intervals are
specified by the boundaries \(0=\kappa_0 < \cdots < \kappa_J = t_{max}\)
and assuming constant baseline hazards \(\lambda_j\) within each
interval the generic PH-model changes to a piecewise exponential model
of the form \begin{equation*}
\boldsymbol{\lambda}(t) = \lambda_j\exp\{\boldsymbol{\eta}\}, \quad \forall \; t\in (\kappa_{j-1}, \kappa_j].
\end{equation*}

Then this formulation is proportional to a Poisson regression of the
events \(\boldsymbol{\delta}_j\) in intervals \(j = 1, \dots, J\) with
expected value \(\text{E}(\boldsymbol{\delta}_j)\) in the sense that
\begin{equation*}
\boldsymbol{\lambda}(t) = \lambda_j\exp\{\boldsymbol{\eta}\} = \frac{\text{E}(\boldsymbol{\delta}_j)}{\exp\{\boldsymbol{o}_j\}}  , \quad \text{where} \quad \text{E}(\boldsymbol{\delta}_j) = \exp\{\log \lambda_j + \boldsymbol{\eta} + \boldsymbol{o}_j\}
\end{equation*} with transformed exposure times
\(\boldsymbol{o}_j = (o_{1j}, \dots, o_{nj})'\) of each individual \(i\)
in each interval \(j\) as offsets (\(\exp\{o_{ij}\} = t_{ij}\))
(Friedman 1982). This further generalises to a piecewise additive mixed
model (PAMM) when the interval-specific log-baseline hazard
\(\log\lambda_j\) is represented as a smooth function of time
\(f_0(t_j)\) instead of a step-function and the predictor
\(\boldsymbol{\eta}\) contains (non-)linear, geographical and/or random
effects (Bender, Groll, and Scheipl 2018).\\
This form of estimation requires the data to be structured differently
than in the conventional way. Table \ref{tab:dataug} gives an example of
this data augmentation and more details can be found in Bender, Groll,
and Scheipl (2018).

(ref:dataaug) Illustration of data augmentation used for applying
Poisson regression. Data augmentation in this toy example was carried
out using \texttt{pammtools} (Bender and Scheipl 2018).

\begin{table}

\caption{\label{tab:dataug}(ref:dataaug)}
\centering
\begin{tabular}[t]{cccccccccccc}

\multicolumn{5}{c}{\makecell[c]{Standard data set for\\proportional hazards approach}} & \multicolumn{1}{c}{ } & \multicolumn{6}{c}{\makecell[c]{Augmented data set for \\piecewise exponential approach}} \\
\cmidrule(l{3pt}r{3pt}){1-5} \cmidrule(l{3pt}r{3pt}){7-12}
$i$ & $\delta_i$ & $T_i$ & $t_i$ & $x_i$ &  & $i$ & $\kappa_{j-1}$ & $\kappa_j$ & $o_j$ & $\delta_j$ & $x_j$\\
\cmidrule(l{3pt}r{3pt}){1-5} \cmidrule(l{3pt}r{3pt}){7-12}
1 & 1 & 0.85 & 0 & 0.83 &  & 1 & 0.0 & 0.30 & -1.20 & 0 & 0.83\\
1 & 1 & 0.85 & 0.3 & -0.28 &  & 1 & 0.3 & 0.40 & -2.30 & 0 & -0.28\\
1 & 1 & 0.85 & 0.6 & -0.36 &  & 1 & 0.4 & 0.60 & -1.61 & 0 & -0.28\\
2 & 0 & 0.58 & 0 & 0.09 & $\qquad \rightarrow \qquad$ & 1 & 0.6 & 0.85 & -1.39 & 1 & -0.36\\
2 & 0 & 0.58 & 0.4 & 2.25 &  & 2 & 0.0 & 0.30 & -1.20 & 0 & 0.09\\
\cmidrule(l{3pt}r{3pt}){1-5}
 &  &  &  &  &  & 2 & 0.3 & 0.40 & -2.30 & 0 & 0.09\\
 &  &  &  &  &  & 2 & 0.4 & 0.60 & -1.71 & 0 & 2.25\\
 \cmidrule(l{3pt}r{3pt}){7-12}
\end{tabular}
\end{table}

\hypertarget{structured-piecewise-additive-joint-models-spajm}{%
\subsection{Structured Piecewise Additive Joint Models
(SPAJM)}\label{structured-piecewise-additive-joint-models-spajm}}

Transferring this counting process representation to the context of
joint models changes the notation thereof to \begin{align}
\boldsymbol{y}(t) & = \boldsymbol{\eta}_\text{l}(t) + \boldsymbol{\eta}_{\text{ls}}(t) + \varepsilon, \quad \varepsilon \sim \text{N}(0, \sigma^2\boldsymbol{I}) \label{rappl:JMPAMMlong}\\
\boldsymbol{\lambda}(t) & = \exp \left\{f_0(t_j) + \boldsymbol{\eta}_\text{s} + \alpha \boldsymbol{\eta}_{\text{ls}} \right\}, \quad \forall \; t\in (\kappa_{j-1}, \kappa_j].\label{rappl:JMPAMMsurv}
\end{align}

The likelihoods then follow the distributions \begin{gather*}
\boldsymbol{y} \mid \boldsymbol{\eta}_{\text{l}}(t), \boldsymbol{\eta}_{\text{ls}}(t) \sim \text{N}(\boldsymbol{\eta}_{\text{l}}(t) + \boldsymbol{\eta}_{\text{ls}}(t), \sigma_\varepsilon^2\boldsymbol{I}) \quad \text{and}\\
\boldsymbol{\delta}_j \mid \boldsymbol{\eta}_{\text{s}}, \boldsymbol{\eta}_{\text{ls}}(t) \sim \text{Poi}(\exp \left\{f_0(t_j) + \boldsymbol{o}_j + \boldsymbol{\eta}_\text{s} + \alpha \boldsymbol{\eta}_{\text{ls}} \right\}) \quad \forall \; t\in (\kappa_{j-1}, \kappa_j]
\end{gather*} for the longitudinal and the time-to-event submodel
respectively.

\hypertarget{posterior-estimation-and-implementation}{%
\subsection{Posterior estimation and
implementation}\label{posterior-estimation-and-implementation}}

Posterior estimation of this model is accomplished via a Markov Chain
Monte Carlo (MCMC) sampler, which in short is a combination of
Gibbs-sampling and a Metropolis-Hastings (MH)-algorithm with iteratively
weighted least squares (IWLS) proposals. The steps of this sampler are
outlined in the following:

\begin{enumerate}
\def\labelenumi{\arabic{enumi}.}
\setcounter{enumi}{-1}
\item
  Initiate starting values for parameter vector
  \(\boldsymbol{\theta}^{[0]} = (\boldsymbol{\theta}^{[0]}_\text{l}; \boldsymbol{\theta}^{[0]}_\text{ls}; \boldsymbol{\theta}^{[0]}_\text{s})'\)
  with\\
  \(\boldsymbol{\theta}^{[0]}_\text{l} = (\boldsymbol{\gamma}^{[0]}_{\text{l}, 1}, \dots, \boldsymbol{\gamma}^{[0]}_{\text{l}, p_\text{l}}, \sigma^{2[0]}_\varepsilon, \sigma^{2[0]}_{\gamma_{\text{l}, 1}}, \dots, \sigma^{2[0]}_{\gamma_{\text{l}, p_\text{l}}})'\)\\
  \(\boldsymbol{\theta}^{[0]}_\text{ls} = (\boldsymbol{\gamma}^{[0]}_{\text{ls}, 1}, \dots, \boldsymbol{\gamma}^{[0]}_{\text{ls}, p_\text{ls}}, \sigma^{2[0]}_{\gamma_{\text{ls}, 1}}, \dots, \sigma^{2[0]}_{\gamma_{\text{ls}, p_\text{ls}}})'\)\\
  \(\boldsymbol{\theta}^{[0]}_\text{s} = (\boldsymbol{\gamma}^{[0]}_{\text{s}, 1}, \dots, \boldsymbol{\gamma}^{[0]}_{\text{s}, p_\text{s}}, \sigma^{2[0]}_{\gamma_{\text{s}, 1}}, \dots, \sigma^{2[0]}_{\gamma_{\text{s}, p_\text{s}}}, \alpha^{[0]}, \sigma^{2[0]}_{\alpha}, \boldsymbol{\gamma}^{[0]}_t, \sigma^{2[0]}_{\gamma_t})'\)\\
  \textcolor{white}{dummytext}\\
  \textbf{For} \(t = {1, \dots, T}\) do
\item
  \textbf{Longitudinal effects: Gibbs-update}\\
  For \(k = 1, \dots, p_\text{l}\) draw
  \(\boldsymbol{\gamma}^{[t]}_{\text{l}, k}\) from
  \(\text{N}\left(\mu^*_{\boldsymbol{\gamma}_{\text{l}, k}}, \Sigma^*_{\boldsymbol{\gamma}_{\text{l}, k}}\right)\)
  with\\
  \[\Sigma^*_{\boldsymbol{\gamma}_{\text{l}, k}} = \left(\frac{1}{\sigma^2_\varepsilon}\boldsymbol{Z}_{l, k}'\boldsymbol{Z}_{l, k} + \frac{1}{\sigma^2_{\boldsymbol{\gamma}_{\text{l}, k}}} \boldsymbol{K}_{\text{l}, k}\right)^{-1} , \qquad \mu^*_{\boldsymbol{\gamma}_{\text{l}, k}}= \Sigma^*_{\boldsymbol{\gamma}_{\text{l}, k}} \left(\frac{1}{\sigma^2_\varepsilon}\left(\boldsymbol{Z}_{l, k}'(\boldsymbol{y} - \boldsymbol{\eta}_\text{l, -k} - \boldsymbol{\eta}_\text{ls})\right) +  \frac{1}{\sigma^2_{\boldsymbol{\gamma}_{\text{l}, k}}} \boldsymbol{K}_{\text{l}, k} \right).\]
  In \(\mu_{\boldsymbol{\gamma}_{\text{l}, k}}\) and
  \(\Sigma_{\boldsymbol{\gamma}_{\text{l}, k}}\) use
  \(\sigma^{2[t-1]}_\varepsilon\),
  \(\sigma^{2[t-1]}_{\gamma_{\text{l}, k}}\),
  \(\boldsymbol{\eta}^{[t-1]}_\text{ls}\) and
  \(\boldsymbol{\eta}^{[t-1]}_{\text{l}, -k} = \boldsymbol{\eta}^{[t-1]}_{\text{l}} - \boldsymbol{\eta}^{[t-1]}_{l,k}\).
\item
  \textbf{Survival effects: IWLS-MH-update}\\
  For \(k = 1, \dots, p_\text{s}\) determine
  \(\boldsymbol{\gamma}^{[t]}_{\text{s}, k}\) as follows:\\
  Draw IWLS proposal \(\boldsymbol{\gamma}_{\text{s}, k}^*\) from
  \(q\left(\boldsymbol{\gamma}_{\text{s}, k}^* \mid \boldsymbol{\gamma}_{\text{s}, k}^{[t-1]}\right) = N\left(\boldsymbol{\mu}_{\boldsymbol{\gamma}_{\text{s}, k}}, \boldsymbol{P}^{-1}_{\boldsymbol{\gamma}_{\text{s}, k}}\right)\)
  with \[ 
  \boldsymbol{P}_{\boldsymbol{\gamma}_{\text{s}, k}} = \boldsymbol{Z}_{\text{s}, k}'\boldsymbol{W}_\text{s}\boldsymbol{Z}_{\text{s}, k} + \frac{1}{\sigma^{2}_{\boldsymbol{\gamma}_{\text{s}, k}}} \boldsymbol{K}_{\boldsymbol{\gamma}_{\text{s}, k}} 
  \quad \text{and} \quad 
  \boldsymbol{\mu}_{\boldsymbol{\gamma}_{\text{s}, k}} = \left(\boldsymbol{P}_{\boldsymbol{\gamma}_{\text{s}, k}}\right)^{-1} \boldsymbol{Z}_{\text{s}, k}'\boldsymbol{W}_\text{s}\left(\tilde{\boldsymbol{y}}_\text{s} - \boldsymbol{\eta}_{\text{s},-k}\right).
  \] In \(\boldsymbol{P}_{\boldsymbol{\gamma}_{\text{s}, k}}\) and
  \(\boldsymbol{\mu}_{\boldsymbol{\gamma}_{\text{s}, k}}\) use
  \(\sigma^{2[t-1]}_{\boldsymbol{\gamma}_{\text{s}, k}}\),
  \(\boldsymbol{\eta}^{[t-1]}_{\text{s}, -k} = \boldsymbol{\eta}^{[t-1]}_{\text{s}} - \boldsymbol{\eta}^{[t-1]}_{s,k}\),
  working weights \(\boldsymbol{W}_\text{s}\) and working observations
  \(\tilde{\boldsymbol{y}_\text{s}}\). The definition of working weights
  and observations is given in appendix \ref{app:JMPAMM:IWLSsurv}.\\
  Accept draw \(\boldsymbol{\gamma}_{\text{s}, k}^*\) with probability
  \[\alpha\left(\boldsymbol{\gamma}_{\text{s}, k}^* \mid \boldsymbol{\gamma}_{\text{s}, k}^{[t]}\right) = \min \left\{\frac{L\left(\boldsymbol{\gamma}_{\text{s}, k}^*\right) \, p\left(\boldsymbol{\gamma}_{\text{s}, k}^*\right) \, q\left(\boldsymbol{\gamma}_{\text{s}, k}^{[t]} \mid \boldsymbol{\gamma}_{\text{s}, k}^*\right)}{L\left(\boldsymbol{\gamma}_{\text{s}, k}^{[t]}\right) \, p\left(\boldsymbol{\gamma}_{\text{s}, k}^{[t]}\right) \, q\left(\boldsymbol{\gamma}_{\text{s}, k}^* \mid \boldsymbol{\gamma}_{\text{s}, k}^{[t]}\right)}, 1 \right\}\]
  with likelihood
  \(L(\boldsymbol{\gamma}_{\text{s}, k}) = p(\boldsymbol{\delta} \mid \boldsymbol{\gamma}_{\text{s}, k}, \cdot)\).
\item
  \textbf{Shared effects: IWLS-MH-update}\\
  For \(k = 1, \dots, p_\text{ls}\) determine
  \(\boldsymbol{\gamma}^{[t]}_{\text{ls}, k}\) as follows:\\
  Draw IWLS proposal \(\boldsymbol{\gamma}_{\text{ls}, k}^*\) from
  \(q\left(\boldsymbol{\gamma}_{\text{ls}, k}^* \mid \boldsymbol{\gamma}_{\text{ls}, k}^{[t-1]}\right) = N\left(\boldsymbol{\mu}_{\boldsymbol{\gamma}_{\text{ls}, k}}, \boldsymbol{P}^{-1}_{\boldsymbol{\gamma}_{\text{ls}, k}}\right)\)
  with \[ 
  \boldsymbol{P}_{\boldsymbol{\gamma}_{\text{ls}, k}} = \boldsymbol{Z}_{\text{ls}, k}'\boldsymbol{W}_\text{ls}\boldsymbol{Z}_{\text{ls}, k} + \frac{1}{\sigma^{2}_{\boldsymbol{\gamma}_{\text{ls}, k}}} \boldsymbol{K}_{\boldsymbol{\gamma}_{\text{ls}, k}} 
  \quad \text{and} \quad 
  \boldsymbol{\mu}_{\boldsymbol{\gamma}_{\text{ls}, k}} = \left(\boldsymbol{P}_{\boldsymbol{\gamma}_{\text{ls}, k}}\right)^{-1} \boldsymbol{Z}_{\text{ls}, k}'\boldsymbol{W}_\text{ls}\left(\tilde{\boldsymbol{y}}_\text{ls} - \boldsymbol{\eta}_{\text{ls},-k}\right).
  \] In \(\boldsymbol{P}_{\boldsymbol{\gamma}_{\text{ls}, k}}\) and
  \(\boldsymbol{\mu}_{\boldsymbol{\gamma}_{\text{ls}, k}}\) use
  \(\sigma^{2[t-1]}_{\boldsymbol{\gamma}_{\text{ls}, k}}\),
  \(\boldsymbol{\eta}^{[t-1]}_{\text{ls}, -k} = \boldsymbol{\eta}^{[t-1]}_{\text{ls}} - \boldsymbol{\eta}^{[t-1]}_{s,k}\),
  working weights \(\boldsymbol{W}_\text{ls}\) and working observations
  \(\tilde{\boldsymbol{y}_\text{ls}}\). The definition of working
  weights and observations is given in appendix
  \ref{app:JMPAMM:IWLSshared}.\\
  Accept draw \(\boldsymbol{\gamma}_{\text{ls}, k}^*\) with probability
  \[\alpha\left(\boldsymbol{\gamma}_{\text{ls}, k}^* \mid \boldsymbol{\gamma}_{\text{ls}, k}^{[t]}\right) = \min \left\{\frac{L\left(\boldsymbol{\gamma}_{\text{ls}, k}^*\right) \, p\left(\boldsymbol{\gamma}_{\text{ls}, k}^*\right) \, q\left(\boldsymbol{\gamma}_{\text{ls}, k}^{[t]} \mid \boldsymbol{\gamma}_{\text{ls}, k}^*\right)}{L\left(\boldsymbol{\gamma}_{\text{ls}, k}^{[t]}\right) \, p\left(\boldsymbol{\gamma}_{\text{ls}, k}^{[t]}\right) \, q\left(\boldsymbol{\gamma}_{\text{ls}, k}^* \mid \boldsymbol{\gamma}_{\text{ls}, k}^{[t]}\right)}, 1 \right\}\]
  with likelihood
  \(L(\boldsymbol{\gamma}_{\text{ls}, k}) = p(\boldsymbol{y} \mid \boldsymbol{\gamma}_{\text{ls}, k}, \cdot) \; p(\boldsymbol{\delta} \mid \boldsymbol{\gamma}_{\text{ls}, k}, \cdot)\).
\item
  \textbf{Update variance parameters: Gibbs-update}\\
\end{enumerate}

\begin{itemize}
\tightlist
\item
  Model variance\\
  Let \(N = \sum_{i = 1}^{n} n_i\) be the total number of longitudinal
  observations as the sum of all observations \(n_i\) per individual
  \(i\) across all individuals \(n\).\\
  Draw \(\sigma^{2[t]}_\varepsilon\) from
  \(\text{IG}(\tilde{a}_0, \tilde{b}_0)\) with\\
  \[\tilde{a}_0 = a_0 + \frac{N}{2}, \qquad \tilde{b}_0 = b_0 + (\boldsymbol{y} - \boldsymbol{\eta}_\text{l} - \boldsymbol{\eta}_\text{ls})'(\boldsymbol{y} - \boldsymbol{\eta}_\text{l} - \boldsymbol{\eta}_\text{ls}).\]\\
  In \(\tilde{a}_0\) and \(\tilde{b}_0\) use
  \(\boldsymbol{\eta}^{[t]}_\text{l}\) and
  \(\boldsymbol{\eta}^{[t]}_\text{ls}\).\\
\item
  Effect variance\\
  For \(k_{\cdot} = 1, \dots, p_{\cdot}\) draw
  \(\sigma^{2[t]}_{\boldsymbol{\gamma}_{k_\cdot}}\) from
  \(\text{IG}(\tilde{a}_{k_{\cdot}}, \tilde{b}_{k_{\cdot}})\) with\\
  \[\tilde{a}_{k_{\cdot}} = a_{k_{\cdot}} + \text{rk}(\boldsymbol{K}_{k_{\cdot}}), \qquad \tilde{b}_{k_{\cdot}} = b_{k_{\cdot}} + \frac{1}{2} \boldsymbol{\gamma}_{\cdot, k_{\cdot}}' \boldsymbol{K}_{k_{\cdot}} \boldsymbol{\gamma}_{\cdot, k_{\cdot}}.\]
  In \(\tilde{a}_{k_{\cdot}}\) and \(\tilde{b}_{k_{\cdot}}\) use
  \(\boldsymbol{\gamma}^{[t]}_{\cdot, k_{\cdot}}\).\\
  The above described algorithm is implemented in the current developer
  version of the statistical software BayesX (Belitz et al. 2022).
\end{itemize}

\hypertarget{simulation-study}{%
\section{Simulation Study}\label{simulation-study}}

With the following simulation study we want to (a) illustrate the
flexibility of the SPAJM with regard to effect specification, (b)
highlight its capability for estimating spatial effects and (c) confirm
its computational advantage by comparing the performance of our approach
to an already existing one. In order to meet intention (a) the simulated
model will be maximally generic, i.e.~include various types of effects
in all possible predictors alongside a non-linear baseline hazard, and
to meet intention (b) the model will comprise a spatial effect. Since
the spatial effect can only be located in one of the predictors for
identifiability reasons we will look at three settings to determine
whether the quality of performance is location specific:

~~\emph{Setting 1} The spatial effect is located in the shared predictor
\(\eta_\text{ls}\),\\
\hspace*{0.333em}\hspace*{0.333em}\emph{Setting 2} it is located in the
survival predictor \(\eta_\text{s}\) and\\
\hspace*{0.333em}\hspace*{0.333em}\emph{Setting 3} in the longitudinal
predictor \(\eta_\text{l}\).

Lastly, to ascertain intention (c) the runtimes of our approach will be
contrasted to an already existing one.\\
In terms of software we will use the BayesX implementation of the SPAJM
and benchmark it against the similarly flexible joint model
implementation of the tensor-product approach using Newton-Raphson
procedures and derivative-based Metropolis-Hastings sampling by Köhler,
Umlauf, et al. (2017) in the \texttt{R} package \texttt{bamlss} (Umlauf
et al. 2021). We will use the current developer version of BayesX
(Belitz et al. 2022) as well as \texttt{bamlss} version 1.1-8 on
\texttt{R}-4.1.2 (R Core Team 2022).

\hypertarget{setup}{%
\subsection{Setup}\label{setup}}

We generate longitudinal measurements \(\boldsymbol{y}(t)\) for
\(n = 200\) individuals over \(n_i = 6\) individual specific, original
time points each in the range of \(t \in (0,1)\) according to the
generic model given in \eqref{rappl:JMPAMMlong} and
\eqref{rappl:JMPAMMsurv} with \(\alpha = -0.3\) and the following
predictors \begin{align*}
\boldsymbol{\eta}_\text{l} & = 0.5\;\boldsymbol{x}_{\text{l}1} + f_{1}(\boldsymbol{x}_{\text{l}2}), \\
\boldsymbol{\eta}_{\text{ls}} &=  0.9\;\boldsymbol{x}_{\text{ls}1} - 0.5\; f_2(\boldsymbol{x}_{\text{ls}2}) - 0.5\;\boldsymbol{x}_{\text{l}3}(t) + 0.4\;\boldsymbol{t} + \boldsymbol{b}_0 + \boldsymbol{b}_1\;\boldsymbol{t} \quad \text{and}\\ 
\boldsymbol{\eta}_\text{s} &= 0.1\;\boldsymbol{x}_{\text{s}1} + 0.5\;f_2(\boldsymbol{x}_{\text{s}2})\\
\end{align*} with the non-linear functions
\(f_1(x) = 0.5\; x + 15\;\phi(2(x-0.2)) - \phi(x + 0.4)\) and
\(f_2(x) = \sin(x)\). All covariates \(\boldsymbol{x}_{\text{ls}\cdot}\)
and \(\boldsymbol{x}_{\text{s}\cdot}\) are simulated as time constant
with the exception of \(\boldsymbol{x}_{\text{ls}3}\), which is
simulated time dependent just like covariates
\(\boldsymbol{x}_{\text{l}\cdot}\), with all
\(\boldsymbol{x}_{\cdot\cdot} \sim U(-1,1)\). Further the model variance
is set to \(\sigma_\varepsilon^2 = 0.5\) and the variances of the random
intercepts and slopes are set to
\(\sigma_{b_0}^2 = \sigma_{b_1}^2 = 2\).\\
True survival times \(T^*_i\) are determined based on a Weibull baseline
hazard function \(\lambda_0(t) = pqt^{q - 1}\) with scale \(p = 0.4\)
and shape \(q = 1.5\). The event times are then set to
\(T_i = \min(T^*_i, 1)\) with event indicator \(\delta_i = 1\) if
\(T^*_i \leq 1\) and \(\delta_i = 0\) otherwise for censored
individuals. For a more realistic censoring scenario we apply in
addition uniform censoring \(U(0,1)\) to 50\% of the censored
individuals.\\
The spatial effect is based on the map of counties in western Germany
available from the \texttt{R} package \texttt{BayesX} and calculated as
\(f_{geo} = \sin(\boldsymbol{c}_x) \cdot \cos(0.5 \; \boldsymbol{c}_y)\)
with \(\boldsymbol{c}_x\) and \(\boldsymbol{c}_y\) being the scaled x-
and y-coordinates respectively of the centroids of each region. The
regions are then randomly distributed across the individuals.\\
For each setting we use \(R=100\) replications. Convergence is achieved
in BayesX by using 70000 iterations per run with a burn-in of 10000 and
a thinning factor of 60 and in \texttt{bamlss} by using 44000 iterations
(54000 with \(f_{geo}\) in \(\eta_{\text{l}}\)) with a burn-in of 4000
and a thinning factor of 40 (50). In order to compare the results of
both implementations we calculate the mean squared error (MSE), bias and
coverage of the 95\%-high density interval (HDI) of the posterior
distribution of each parameter and compare runtimes between BayesX and
\texttt{bamlss}.

\hypertarget{results}{%
\subsection{Results}\label{results}}

\begin{figure}

{\centering \includegraphics{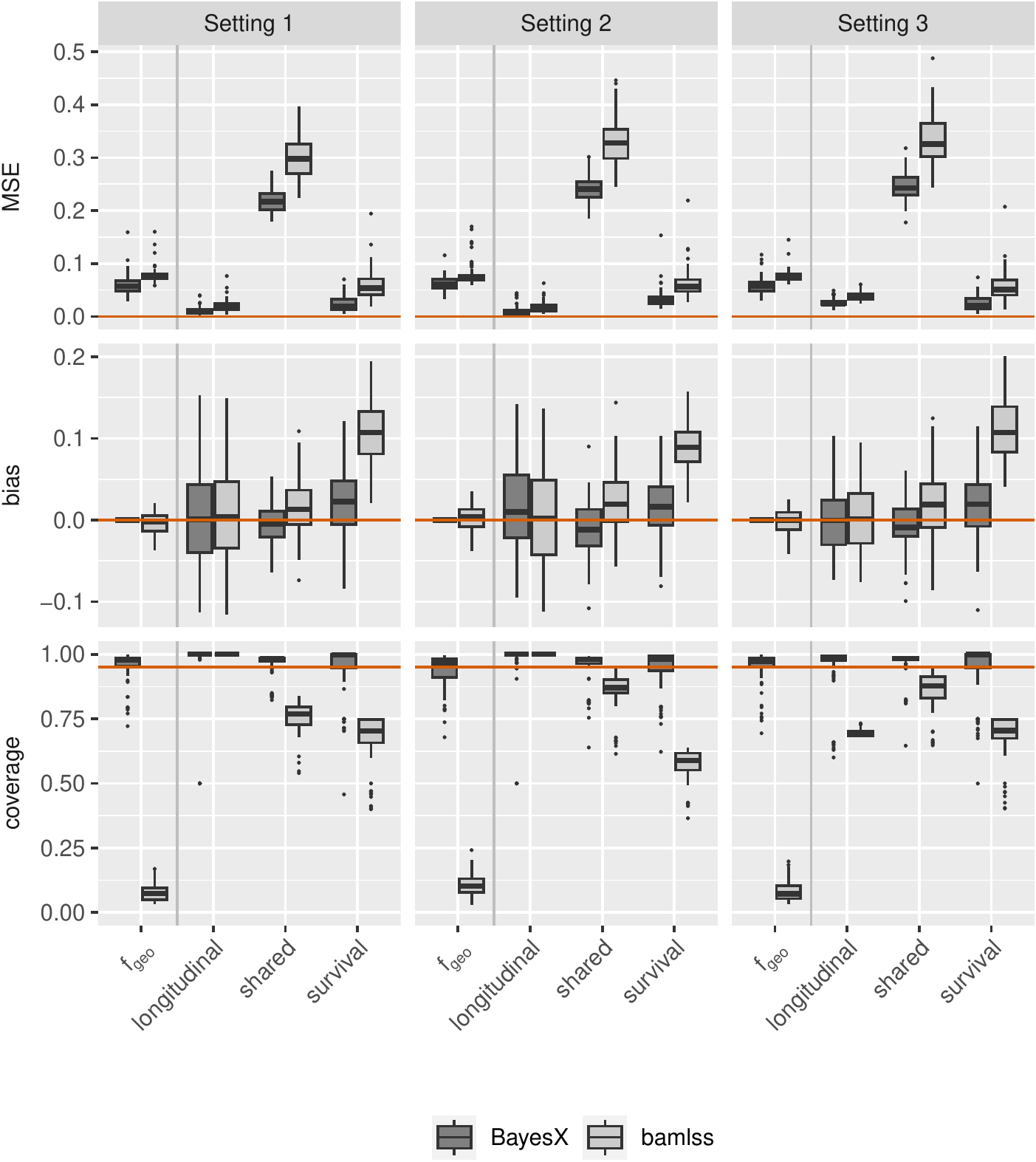} 

}

\caption{Boxplots of mean squared error (MSE), bias and 95\%-coverage for the geographic effect $f_{geo}$ as well as the predictors per method and simulation setting (Setting 1 - $f_{geo}$ in $\eta_{ls}$, Setting 2 - $f_{geo}$ in $\eta_{s}$, Setting 3 - $f_{geo}$ in $\eta_{l}$). The orange horizontal line marks the reference value of each statistic.}\label{fig:simres}
\end{figure}

\begin{figure}
\centering
\includegraphics{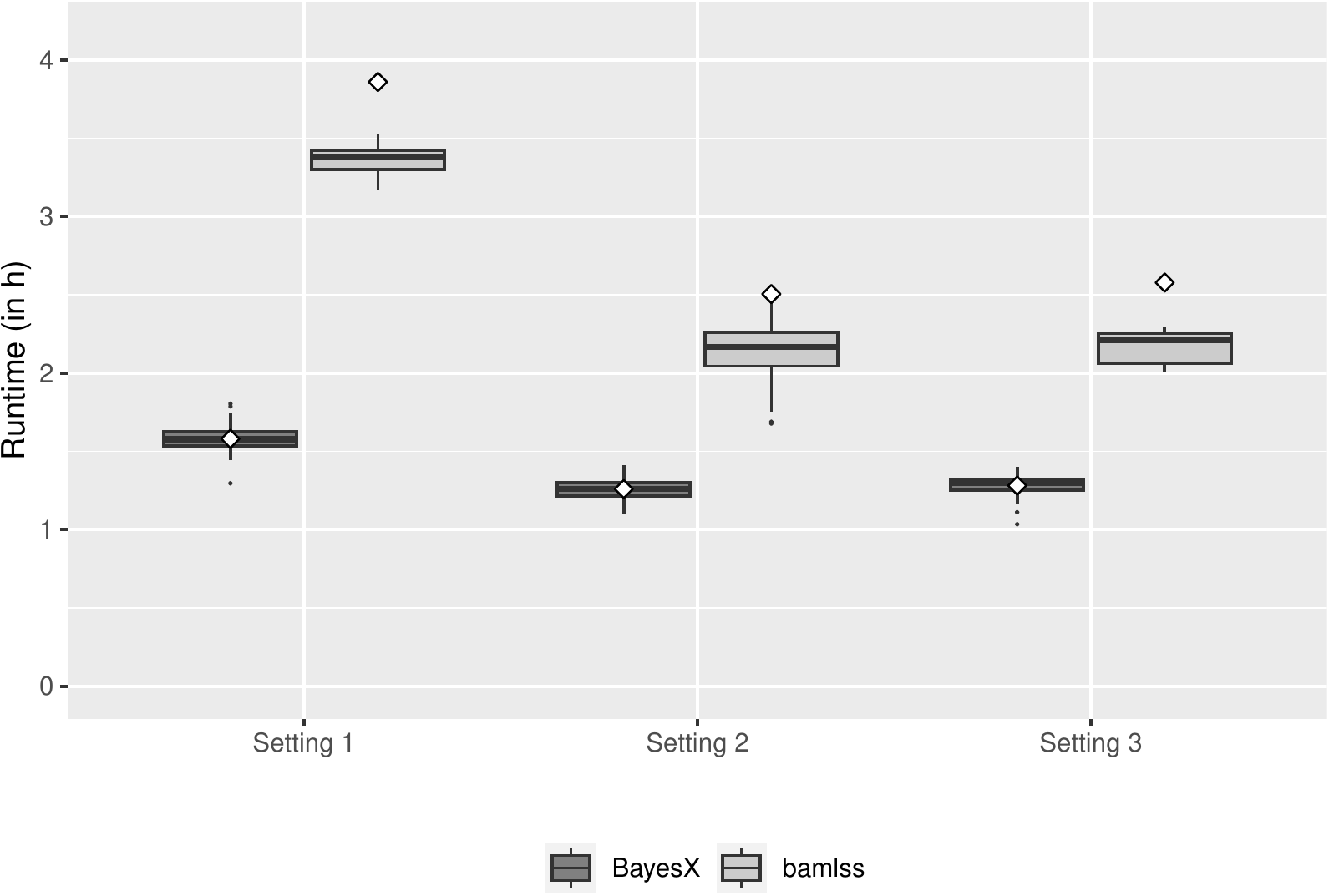}
\caption{Boxplots of runtimes in hours per method and simulation
setting. For readability reasons plots are clipped, therefore extreme
outliers are excluded. Diamonds represent respective means. BayesX has
visibly lower runtimes than \texttt{bamlss}. With both methods Setting 1
(\(f_{geo}\) in \(\eta_{ls}\)) is the most time consuming, while Setting
2 (\(f_{geo}\) in \(\eta_{s}\)) and 3 (\(f_{geo}\) in \(\eta_{l}\)) are
faster and take equally long.\label{fig:runtimes}}
\end{figure}

The outcome of the estimation performance of the simulation study can be
found in Figure \ref{fig:simres} and the computational performance is
illustrated in Figure \ref{fig:runtimes}. The summarized results in
Figure \ref{fig:simres} already make it clear that a joint model with a
piecewise additive formulation of the survival submodel is is equal in
terms of effect estimation to its established PH counterpart given the
small MSE and bias values as well as the high coverage rates. Detailed
results of the individual effects can be found in the appendix in Figure
\ref{fig:appres}, which confirm this high level impression. Both methods
exhibit the largest deviation from the true data in the shared
predictors \(\eta_{\text{ls}}\). In terms of estimation any effect in
this predictor belongs to the most demanding to estimate, as the
corresponding likelihood features both model parts. Thus the larger bias
here is to be expected. Furthermore, it quickly becomes clear that
BayesX outperforms \texttt{bamlss} in the estimation results of the
shared predictors \(\eta_{\text{ls}}\) and the survival predictors
\(\eta_\text{s}\). The reason for the performance of \texttt{bamlss} in
the shared predictors \(\eta_{\text{ls}}\) is due to the random effects,
which can be seen from the more detailed Figure \ref{fig:appres} in the
Appendix. Their estimates remain rather small, which is why their high
density intervals do not cover the simulated (true) random effects
\(\boldsymbol{b}_0\) and \(\boldsymbol{b}_1\), which in turn affects the
overall results for the shared predictor \(\eta_{\text{ls}}\). Similarly
the survival predictors \(\eta_{\text{s}}\) perform rather weak with
\texttt{bamlss}, which is mainly due to the rather large bias in the
association \(\alpha\) (see Figure \ref{fig:appres}). The estimation
procedure implemented in \texttt{bamlss} is in fact tailored to identify
advanced association structures in joint models, which is why the bias
in \(\alpha\) is highly likely a result of the underestimation of the
random effects. Only in the estimation of the longitudinal predictor
\(\eta_{\text{l}}\) did \texttt{bamlss} surpass BayesX, which is
interesting, since the formulation implemented in \texttt{bamlss} does
not extend to longitudinal-only-predictors. It assumes
\(\eta_{\text{l}}\) to be a part of \(\eta_{\text{ls}}\), but since the
data of \(\eta_{\text{l}}\) is simulated such that it is not associated
with the survival part of the model, the results of \(\eta_{\text{l}}\)
under \texttt{bamlss} are more precise than those of
\(\eta_{\text{ls}}\). Figure \ref{fig:simres} further demonstrates the
capability of both methods to estimate spatial effects, while it also
shows the indifference to the position (in \(\eta_\text{ls}\),
\(\eta_\text{s}\) or \(\eta_\text{l}\)) of the spatial effect within the
model. First of all, the figure indicates a stable performance of the
spatial effect \(f_{geo}\) in both implementations independent of the
predictor it belongs to. Secondly, also the other predictors remain very
stable in their performance regardless of the simulation setting. If the
position of the geographical effect \(f_{geo}\) mattered, it would not
just show in the estimation accuracy of the effect itself, but it would
also affect the effects in other parts of the model, which is not the
case here. Again the reason for the \texttt{bamlss} results are similar
to before. The estimation results for \(f_{geo}\) exhibit the same
behaviour as for the random effects: They remain surprisingly small,
resulting in a larger bias and thus only achieving a rather low
coverage.\\
Lastly, in terms of computational cost the piecewise additive approach
in BayesX has an advantage over the PH approach in \texttt{bamlss} with
lower runtimes (see Figure \ref{fig:runtimes}). With both methods
Setting 1 with the geographic effect \(f_{geo}\) in the shared predictor
\(\eta_{ls}\) is the most time consuming. But this is also the most
complex setting in terms of estimation, therefore, the increased runtime
is not surprising. Setting 2 with \(f_{geo}\) in the survival predictor
\(\eta_{s}\) and Setting 3 with \(f_{geo}\) in the longitudinal
predictor \(\eta_{l}\) are less complex from an estimation perspective,
which is also evident in the short runtimes. More detailed descriptive
statistics on the runtimes can be found in the appendix in Table
\ref{tab:app_runtimes}.

\hypertarget{physical-functioning-after-a-caesura}{%
\section{Physical Functioning after a
Caesura}\label{physical-functioning-after-a-caesura}}

In 2015 the World Health Organsiation (WHO) concluded in their ``World
Report on Ageing and Health'' that the physical capacity dimension of
``Healthy Ageing'' still suffers from a lack of understanding. Physical
capacity can be measured as functional health (aka physical
functioning), which decreases naturally over time until death. However,
certain physiological events have the power to alter the trajectory of
an individual's functional health both in a negative and positive way,
among them heart attacks, strokes or diagnoses of cancer (caesura, WHO
2015). While the longitudinal modelling of these trajectories is already
of interest, the trajectories themselves influence an individual's
survival time. Therefore, a joint model is appropriate to capture both
these aspects of the data.\\
To examine the development of physical functioning after a caesura in
Germany we will resort to the German Ageing Survey (DEAS), which aims at
studying the second half of life with people between 40 and 85 years old
and living in Germany being eligible for study participation. The DEAS
has collected information on physical functioning from a SF-36 survey,
health conditions qualifying as caesurae, terminal dates and a multitude
of other variables, which might help explain the development of physical
functioning after a caesura, over the course of seven waves (1996, 2002,
2008, 2011, 2014, 2017, 2021) (Klaus and Engstler 2017; Engstler,
Hameister, and Schrader 2014).\\
Our analysis will focus on data from waves 2008 to 2021 with originally
6622 participants, of which 750 suffered from a heart attack or stroke
i.e.~a cardiovascular caesura, during their panel participation. Single
observations, cases with missing data and caesurae with onset prior to
the participant's entry into the panel were excluded from the analysis.
For the remaining 636 the time of onset of the caesura was set to
coincide with the interview date, in which the caesura was first
reported, since the exact onset date of the caesurae is not collected.
Out of 636 participants 79 (12.4\%) died.\\
As explanatory variables for the trajectory of functional health
(\texttt{sf36}) we consider time (\(\boldsymbol{t}\)), gender
(\texttt{gender}), the age of onset (\texttt{aoo}) of the caesura as
well as living location of the participant on the level of European
Nomenclature of Territorial Units for Statistics (NUTS) 2. In order to
avoid re-identification of participants few regions had to be combined
leaving now 33 regions of the original 36. The continuous and strictly
positive variables SF-36 \texttt{sf36}, age of onset \texttt{aoo} and
time \(\boldsymbol{t}\) are scaled to the domain \((0, 1)\). We then
consider the model \begin{align*}
    \texttt{sf36}_i(t) & = \beta_0 + \boldsymbol{\eta}_{\text{ls}i} +\varepsilon_{i}(t)\\
    \lambda(t) &= \exp\{f_0(t) + \alpha \boldsymbol{\eta}_{\text{ls}i}\}\\
    \boldsymbol{\eta}_{\text{ls}i}&=\beta_1\texttt{gender}_i + f(\texttt{aoo}_i) + f_{geo}(\texttt{NUTS2}_i) + \beta_t \boldsymbol{t} + b_{0i} + b_{1i}\boldsymbol{t}
\end{align*} and estimate it with \texttt{BayesX} and \texttt{bamlss}.

\begin{table}[!h]

\caption{\label{tab:datlinres}BayesX estimates of linear effects of physical functioning after a caesura.}
\centering
\begin{tabular}[t]{lcc}
\toprule
 & posterior mean & 95\%-HDI\\
\midrule
$\beta_0$ & 0.848 & {}[0.827, 0.871]\\
gender & -0.091 & {}[-0.123, -0.059]\\
$t$ & -0.247 & {}[-0.277, -0.219]\\
$\alpha$ & -3.381 & {}[-4.329, -2.432]\\
\bottomrule
\end{tabular}
\end{table}

\begin{figure}
\centering
\includegraphics{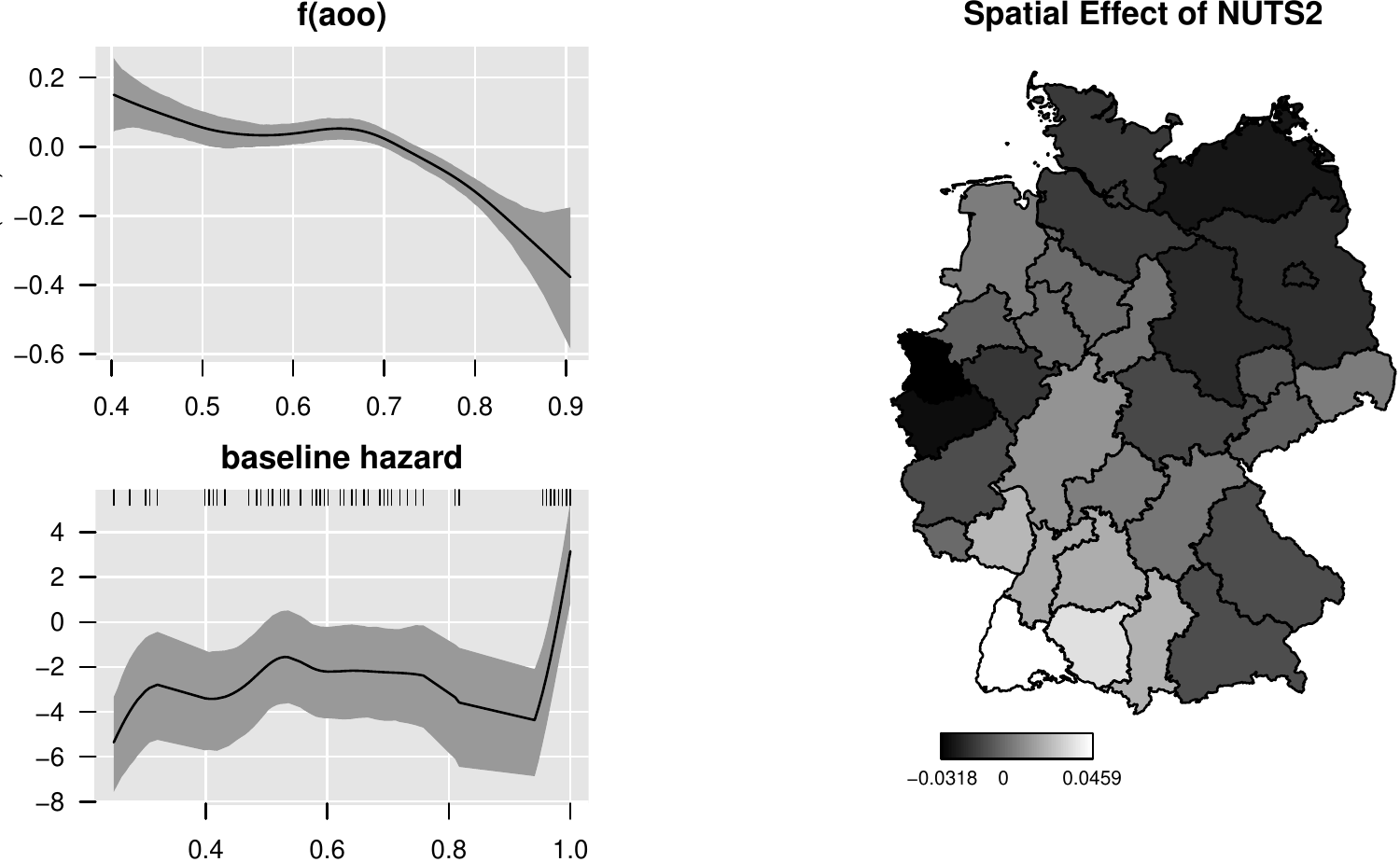}
\caption{BayesX estimates of the smooth effect of the age of onset
\texttt{aoo} and the geographical location on physical functioning as
well as the estimated baseline hazard of the model.
\label{fig:datnlinres}}
\end{figure}

With 79 events out of 636 individuals the survival data is unbalanced
and presents a situation that - in a standard survival analysis setting
- would already prove difficult to estimate. \texttt{bamlss} proved to
react sensitive to this imbalance, which is why we only present BayesX
results here. For the linear effects they can be found in Table
\ref{tab:datlinres} and for the non-linear effects in Figure
\ref{fig:datnlinres}.\\
None of the linear effects includes zero in their HDI, thus they are
significantly different from zero. The association \(\alpha\) of both
model parts is negative meaning that a lower level of the modelled
trajectory of physical functioning \texttt{sf36} translates to a higher
probability of experiencing an event.\\
The intercept can be interpreted as a male individual at scaled age of
0.402 (i.e.~an unscaled age of 40.2) years old at onset of the caesura
can be expected to have an average scaled SF-36 level \texttt{sf36} of
0.848 {[}0.827, 0.871{]}. For women this reduces on average by -0.091
{[}-0.123, -0.059{]}. Every scaled month after the caesura further
reduces the level of \texttt{sf36} by -0.247 {[}-0.277, -0.219{]}. The
age of onset \texttt{aoo} has in general a decreasing effect on
\texttt{sf36} (upper left panel Figure \ref{fig:datnlinres}). Though it
needs to be pointed out that before the scaled age of 0.55 (55 years
old) the effect is positive, i.e.~it increases the level of
\texttt{sf36} thus slowing down the natural decline of physical
functioning, while for an \texttt{aoo} between roughly 0.55 and 0.7 (55
to 70 years old) the effect is constant around zero, i.e.~it is
negligible, and a caesura after an \texttt{aoo} of 0.7 (70 years old)
has a negative effect on \texttt{sf36} translating to an accelerated
decline of physical functioning. In terms of living location there is a
South-West against North-East (and Mid-West) divide (right panel Figure
\ref{fig:datnlinres}). People in the North-Eastern part of Germany
especially in the area of Mecklenburg-Pommerania, Brandenburg and
Saxony-Anhalt as well as those of the Western parts in the Dusseldorf
and Cologne regions see a negative effect on their level of
\texttt{sf36}. Those living in the South-Western part especially in
South-West Baden-Wurttemberg (Black Forest region) see an increasing
effect on their \texttt{sf36} level. Given that the association is
negative this means that the probability for an event is decreased most
for people from the South West of Germany and increased most for those
in the North-East and Mid-West. What these two areas have in common is
that they comprise the most and least densely populated areas in
Germany. This might be a starting point for further research to
investigate what exactly triggers the effect to take this particular
shape, since the living location in this example can be interpreted as a
proxy for other variables that have not been included in the model.\\
The baseline hazard is almost linear over time (lower left panel Figure
\ref{fig:datnlinres}), thus the risk of experiencing an event is roughly
the same at all times throughout the study.

\hypertarget{conclusion-and-discussion}{%
\section{Conclusion and Discussion}\label{conclusion-and-discussion}}

The focus of this article has been on proposing a piecewise additive
joint model for longitudinal and time-to-event data allowing for
spatial, (non-)-linear and random effects to be included as well as
estimation of the baseline hazard without any assumptions about its
distributional form. In a simulation study comprising (non-)linear as
well as a spatial effect it became evident that the piecewise additive
approach yields results similar or better to the equally flexibly
\texttt{bamlss}-methodology for joint models in \texttt{R} and that this
performance is high independent of the position of the spatial effect.
This method was illustrated by an example of the development of physical
functioning after a caesura in people in their second half of life.\\
The concept of piecewise additive joint models has not just proven its
accuracy in estimating complex effects, but also its ability in handling
unbalanced data in terms of availability of event observations.\\
Applying the piecewise additive approach requires augmenting data, which
is part of the time-to-event process. This augmentation artificially
increases the size of the data set and when the original data is large,
it can lead to longer runtimes. In our experience this is, however,
seldom the case. Furthermore, this method could also be combined with
other models in the longitudinal part of the model such as quantile
regression, a location-scale model or multiple longitudinal outcomes in
a multivariate joint model. Also, Bayesian variable or effect selection
in this type of joint model could be investigated since very few methods
for variable selection in joint models exist yet.

\textbf{Acknowledgement}: Elisabeth Bergherr gratefully acknowledges
funding from the Deutsche Forschungsgemeinschaft (DFG, German Research
Foundation), grant WA 4249/2-1.

\hypertarget{references}{%
\section*{References}\label{references}}
\addcontentsline{toc}{section}{References}

\hypertarget{refs}{}
\begin{CSLReferences}{1}{0}
\leavevmode\vadjust pre{\hypertarget{ref-Alsefri.2020}{}}%
Alsefri, Maha, Maria Sudell, Marta García-Fiñana, and Ruwanthi
Kolamunnage-Dona. 2020. {``Bayesian Joint Modelling of Longitudinal and
Time to Event Data: A Methodological Review.''} \emph{BMC Medical
Research Methodology} 20 (1): 1--17.

\leavevmode\vadjust pre{\hypertarget{ref-Andrinopoulou.2014}{}}%
Andrinopoulou, Eleni-Rosalina, Dimitris Rizopoulos, Johanna J. M.
Takkenberg, and Emmanuel Lesaffre. 2014. {``Joint Modeling of Two
Longitudinal Outcomes and Competing Risk Data.''} \emph{Statistics in
Medicine} 33 (18): 3167--78.
https://doi.org/\url{https://doi.org/10.1002/sim.6158}.

\leavevmode\vadjust pre{\hypertarget{ref-Barrett.2019}{}}%
Barrett, Jessica K, Raphael Huille, Richard Parker, Yuichiro Yano, and
Michael Griswold. 2019. {``Estimating the Association Between Blood
Pressure Variability and Cardiovascular Disease: An Application Using
the ARIC Study.''} \emph{Statistics in Medicine} 38 (10): 1855--68.

\leavevmode\vadjust pre{\hypertarget{ref-BayesX}{}}%
Belitz, Christiane, Andreas Brezger, Thomas Kneib, Stefan Lang, and
Nikolaus Umlauf. 2022. \emph{{BayesX}: Software for {B}ayesian Inference
in Structured Additive Regression Models}.
\url{https://www.uni-goettingen.de/de/bayesx/550513.html}.

\leavevmode\vadjust pre{\hypertarget{ref-Bender.2018b}{}}%
Bender, Andreas, Andreas Groll, and Fabian Scheipl. 2018. {``{A
generalized additive model approach to time-to-event analysis}.''}
\emph{Statistical Modelling}.
\url{https://journals.sagepub.com/doi/10.1177/1471082X17748083}.

\leavevmode\vadjust pre{\hypertarget{ref-Bender.2018a}{}}%
Bender, Andreas, and Fabian Scheipl. 2018. {``{pammtools: Piece-wise
exponential Additive Mixed Modeling tools}.''} \emph{arXiv:1806.01042
{[}Stat{]}}. \url{https://arxiv.org/abs/1806.01042}.

\leavevmode\vadjust pre{\hypertarget{ref-Blanche.2015}{}}%
Blanche, Paul, Cécile Proust-Lima, Lucie Loubère, Claudine Berr,
Jean-François Dartigues, and Hélène Jacqmin-Gadda. 2015. {``Quantifying
and Comparing Dynamic Predictive Accuracy of Joint Models for
Longitudinal Marker and Time-to-Event in Presence of Censoring and
Competing Risks.''} \emph{Biometrics} 71 (1): 102--13.
https://doi.org/\url{https://doi.org/10.1111/biom.12232}.

\leavevmode\vadjust pre{\hypertarget{ref-Crowther.2013}{}}%
Crowther, Michael J., Keith R. Abrams, and Paul C. Lambert. 2013.
{``Joint Modeling of Longitudinal and Survival Data.''} \emph{The Stata
Journal} 13 (1): 165--84.
\url{https://doi.org/10.1177/1536867X1301300112}.

\leavevmode\vadjust pre{\hypertarget{ref-Engstler.2014}{}}%
Engstler, Heribert, Nicole Hameister, and Sophie Schrader. 2014. {``User
Manual {DEAS SUF} 2014.''} DZA German Centre of Gerontology.

\leavevmode\vadjust pre{\hypertarget{ref-Faucett.1998}{}}%
Faucett, Cheryl L., Nathaniel Schenker, and Robert M. Elashoff. 1998.
{``Analysis of Censored Survival Data with Intermittently Observed
Time-Dependent Binary Covariates.''} \emph{Journal of the American
Statistical Association} 93 (442): 427--37.
\url{https://doi.org/10.1080/01621459.1998.10473692}.

\leavevmode\vadjust pre{\hypertarget{ref-Faucett.1996}{}}%
Faucett, Cheryl L., and Duncan C. Thomas. 1996. {``Simultaneously
Modelling Censored Survival Data and Repeatedly Measured Covariates: A
{G}ibbs Sampling Approach.''} \emph{Statistics in Medicine} 15 (15):
1663--85.
\url{https://doi.org/10.1002/(SICI)1097-0258(19960815)15:15\%3C1663::AID-SIM294\%3E3.0.CO;2-1}.

\leavevmode\vadjust pre{\hypertarget{ref-Friedmann.1982}{}}%
Friedman, Michael. 1982. {``{Piecewise Exponential Models for Survival
Data with Covariates}.''} \emph{The Annals of Statistics} 10 (1):
101--13. \url{https://doi.org/10.1214/aos/1176345693}.

\leavevmode\vadjust pre{\hypertarget{ref-Griesbach.2021}{}}%
Griesbach, Colin, Andreas Groll, and Elisabeth Bergherr. 2021. {``Joint
Modelling Approaches to Survival Analysis via {Likelihood-Based}
Boosting Techniques.''} \emph{Computational and Mathematical Methods in
Medicine} 2021 (November): 4384035.

\leavevmode\vadjust pre{\hypertarget{ref-Henderson.2000}{}}%
Henderson, Robin, Peter Diggle, and Angela Dobson. 2000. {``Joint
Modelling of Longitudinal Measurements and Event Time Data.''}
\emph{Biostatistics (Oxford, England)} 1 (4): 465--80.

\leavevmode\vadjust pre{\hypertarget{ref-Hickey.2018}{}}%
Hickey, Graeme L., Pete Philipson, Andrea Jorgensen, and Ruwanthi
Kolamunnage-Dona. 2018. {``Joine{RML}: A Joint Model and Software
Package for Time-to-Event and Multivariate Longitudinal Outcomes.''}
\emph{BMC Medical Research Methodology} 18 (1): 50.
\url{https://doi.org/10.1186/s12874-018-0502-1}.

\leavevmode\vadjust pre{\hypertarget{ref-Huang.2011}{}}%
Huang, Xin, Gang Li, Robert M Elashoff, and Jianxin Pan. 2011. {``A
General Joint Model for Longitudinal Measurements and Competing Risks
Survival Data with Heterogeneous Random Effects.''} \emph{Lifetime Data
Analysis} 17 (1): 80--100.

\leavevmode\vadjust pre{\hypertarget{ref-Huang.2016}{}}%
Huang, Yangxin, and Jiaqing Chen. 2016. {``Bayesian Quantile
Regression-Based Nonlinear Mixed-Effects Joint Models for Time-to-Event
and Longitudinal Data with Multiple Features.''} \emph{Statistics in
Medicine} 35 (30): 5666--85.
https://doi.org/\url{https://doi.org/10.1002/sim.7092}.

\leavevmode\vadjust pre{\hypertarget{ref-Jacqmin.2010}{}}%
Jacqmin-Gadda, Hélène, Cécile Proust-Lima, Jeremy M. G. Taylor, and
Daniel Commenges. 2010. {``Score Test for Conditional Independence
Between Longitudinal Outcome and Time to Event Given the Classes in the
Joint Latent Class Model.''} \emph{Biometrics} 66 (1): 11--19.
https://doi.org/\url{https://doi.org/10.1111/j.1541-0420.2009.01234.x}.

\leavevmode\vadjust pre{\hypertarget{ref-Klaus.2017}{}}%
Klaus, Daniela, and Heribert Engstler. 2017. {``{D}aten Und {M}ethoden
Des {D}eutschen {A}lterssurveys.''} In \emph{Altern Im Wandel}, edited
by Katharina Mahne, Julia Katharina Wolff, Julia Simonson, and Clemens
Tesch-Römer, 29--45. Wiesbaden; s.l.: {Springer Fachmedien Wiesbaden}.
\url{https://doi.org/10.1007/978-3-658-12502-8/_2}.

\leavevmode\vadjust pre{\hypertarget{ref-Koehler.2017c}{}}%
Köhler, Meike, Andreas Beyerlein, Kendra Vehik, Sonja Greven, Nikolaus
Umlauf, Åke Lernmark, William A Hagopian, et al. 2017. {``Joint Modeling
of Longitudinal Autoantibody Patterns and Progression to Type 1
Diabetes: Results from the TEDDY Study.''} \emph{Acta Diabetologica} 54
(11): 1009--17.

\leavevmode\vadjust pre{\hypertarget{ref-Koehler.2017a}{}}%
Köhler, Meike, Nikolaus Umlauf, Andreas Beyerlein, Christiane Winkler,
Anette-Gabriele Ziegler, and Sonja Greven. 2017. {``{Flexible Bayesian
additive joint models with an application to type 1 diabetes
research}.''} \emph{Biometrical Journal} 59 (6): 1144--65.
https://doi.org/\url{https://doi.org/10.1002/bimj.201600224}.

\leavevmode\vadjust pre{\hypertarget{ref-Koehler.2017b}{}}%
Köhler, Meike, Nikolaus Umlauf, and Sonja Greven. 2018. {``{Nonlinear
association structures in flexible Bayesian additive joint models}.''}
\emph{Statistics in Medicine} 37 (30): 4771--88.
https://doi.org/\url{https://doi.org/10.1002/sim.7967}.

\leavevmode\vadjust pre{\hypertarget{ref-Lin.2002}{}}%
Lin, Haiqun, Charles E McCulloch, and Susan T Mayne. 2002. {``Maximum
Likelihood Estimation in the Joint Analysis of Time-to-Event and
Multiple Longitudinal Variables.''} \emph{Statistics in Medicine} 21
(16): 2369--82.

\leavevmode\vadjust pre{\hypertarget{ref-Martins.2016}{}}%
Martins, Rui, Giovani L. Silva, and Valeska Andreozzi. 2016.
{``{Bayesian joint modeling of longitudinal and spatial survival AIDS
data}.''} \emph{Statistics in Medicine} 35 (19): 3368--84.
https://doi.org/\url{https://doi.org/10.1002/sim.6937}.

\leavevmode\vadjust pre{\hypertarget{ref-Martins.2017}{}}%
---------. 2017. {``{Joint analysis of longitudinal and survival AIDS
data with a spatial fraction of long-term survivors: A Bayesian
approach}.''} \emph{Biometrical Journal} 59 (6): 1166--83.
https://doi.org/\url{https://doi.org/10.1002/bimj.201600159}.

\leavevmode\vadjust pre{\hypertarget{ref-Mauff.2020}{}}%
Mauff, Katya, Ewout Steyerberg, Isabella Kardys, Eric Boersma, and
Dimitris Rizopoulos. 2020. {``Joint Models with Multiple Longitudinal
Outcomes and a Time-to-Event Outcome: A Corrected Two-Stage Approach.''}
\emph{Statistics and Computing} 30: 999--1014.

\leavevmode\vadjust pre{\hypertarget{ref-R.2022}{}}%
R Core Team. 2022. \emph{R: A Language and Environment for Statistical
Computing}. Vienna, Austria: R Foundation for Statistical Computing.
\url{https://www.R-project.org/}.

\leavevmode\vadjust pre{\hypertarget{ref-Rappl.2022}{}}%
Rappl, Anja, Andreas Mayr, and Elisabeth Waldmann. 2022. {``More Than
One Way: Exploring the Capabilities of Different Estimation Approaches
to Joint Models for Longitudinal and Time-to-Event Outcomes.''}
\emph{The International Journal of Biostatistics} 18 (1): 127--49.
\url{https://doi.org/doi:10.1515/ijb-2020-0067}.

\leavevmode\vadjust pre{\hypertarget{ref-Rizopoulos.2010}{}}%
Rizopoulos, Dimitris. 2010. {``JM: An r Package for the Joint Modelling
of Longitudinal and Time-to-Event Data.''} \emph{Journal of Statistical
Software} 35 (9). \url{https://doi.org/10.18637/jss.v035.i09}.

\leavevmode\vadjust pre{\hypertarget{ref-Rizopoulos.2011a}{}}%
---------. 2011. {``Dynamic Predictions and Prospective Accuracy in
Joint Models for Longitudinal and Time-to-Event Data.''}
\emph{Biometrics} 67 (3): 819--29.
\url{https://doi.org/10.1111/j.1541-0420.2010.01546.x}.

\leavevmode\vadjust pre{\hypertarget{ref-Rizopoulos.2016}{}}%
---------. 2016. {``The {R} Package {JM}bayes for Fitting Joint Models
for Longitudinal and Time-to-Event Data Using {MCMC}.''} \emph{Journal
of Statistical Software} 72 (7).
\url{https://doi.org/10.18637/jss.v072.i07}.

\leavevmode\vadjust pre{\hypertarget{ref-Rizopoulos.2011b}{}}%
Rizopoulos, Dimitris, and Pulak Ghosh. 2011. {``A Bayesian
Semiparametric Multivariate Joint Model for Multiple Longitudinal
Outcomes and a Time-to-Event.''} \emph{Statistics in Medicine} 30 (12):
1366--80. https://doi.org/\url{https://doi.org/10.1002/sim.4205}.

\leavevmode\vadjust pre{\hypertarget{ref-Rizopoulos.2008}{}}%
Rizopoulos, Dimitris, Geert Verbeke, Emmanuel Lesaffre, and Yves
Vanrenterghem. 2008. {``A Two-Part Joint Model for the Analysis of
Survival and Longitudinal Binary Data with Excess Zeros.''}
\emph{Biometrics} 64 (2): 611--19.
https://doi.org/\url{https://doi.org/10.1111/j.1541-0420.2007.00894.x}.

\leavevmode\vadjust pre{\hypertarget{ref-Tseng.2005}{}}%
Tseng, Yi-Kuan, Fushing Hsieh, and Jane-Ling Wang. 2005. {``{Joint
modelling of accelerated failure time and longitudinal data}.''}
\emph{Biometrika} 92 (3): 587--603.
\url{https://doi.org/10.1093/biomet/92.3.587}.

\leavevmode\vadjust pre{\hypertarget{ref-Tsiatis.2004}{}}%
Tsiatis, Anastasios A., and Marie Davidian. 2004. {``Joint Modeling of
Longitudinal and Time-to-Event Data: An Overview.''} \emph{Statistica
Sinica} 14 (3): 809--34.

\leavevmode\vadjust pre{\hypertarget{ref-Umlauf.2021}{}}%
Umlauf, Nikolaus, Nadja Klein, Thorsten Simon, and Achim Zeileis. 2021.
{``{bamlss}: A {L}ego Toolbox for Flexible {B}ayesian Regression (and
Beyond).''} \emph{Journal of Statistical Software} 100 (4): 1--53.
\url{https://doi.org/10.18637/jss.v100.i04}.

\leavevmode\vadjust pre{\hypertarget{ref-Viviani.2014}{}}%
Viviani, Sara, Marco Alfó, and Dimitris Rizopoulos. 2014. {``Generalized
Linear Mixed Joint Model for Longitudinal and Survival Outcomes.''}
\emph{Statistics and Computing} 24 (3): 417--27.

\leavevmode\vadjust pre{\hypertarget{ref-Waldmann.2017}{}}%
Waldmann, Elisabeth, David Taylor-Robinson, Nadja Klein, Thomas Kneib,
Tania Pressler, Matthias Schmid, and Andreas Mayr. 2017. {``Boosting
Joint Models for Longitudinal and Time-to-Event Data.''}
\emph{Biometrical Journal} 59 (6): 1104--21.
https://doi.org/\url{https://doi.org/10.1002/bimj.201600158}.

\leavevmode\vadjust pre{\hypertarget{ref-WHO.2015}{}}%
WHO. 2015. \emph{World Report on Ageing and Health}. Geneva: World
Health Organisation; WHO.

\leavevmode\vadjust pre{\hypertarget{ref-Wulfsohn.1997}{}}%
Wulfsohn, Michael S., and Anastasios A. Tsiatis. 1997. {``A Joint Model
for Survival and Longitudinal Data Measured with Error.''}
\emph{Biometrics} 53 (1): 330. \url{https://doi.org/10.2307/2533118}.

\leavevmode\vadjust pre{\hypertarget{ref-Yuen.2016}{}}%
Yuen, Hok Pan, and Andrew Mackinnon. 2016. {``Performance of Joint
Modelling of Time-to-Event Data with Time-Dependent Predictors: An
Assessment Based on Transition to Psychosis Data.''} \emph{PeerJ} 4:
e2582. \url{https://doi.org/10.7717/peerj.2582}.

\leavevmode\vadjust pre{\hypertarget{ref-Zhang.2019}{}}%
Zhang, Hanze, Yangxin Huang, Wei Wang, Henian Chen, and Barbara
Langland-Orban. 2019. {``Bayesian Quantile Regression-Based Partially
Linear Mixed-Effects Joint Models for Longitudinal Data with Multiple
Features.''} \emph{Statistical Methods in Medical Research} 28 (2):
569--88. \url{https://doi.org/10.1177/0962280217730852}.

\end{CSLReferences}

\newpage

\hypertarget{appendix}{%
\section*{Appendix}\label{appendix}}
\addcontentsline{toc}{section}{Appendix}

\appendix

\hypertarget{derivation-of-the-full-conditional-and-iwls-proposal-distributions-used-for-posterior-estimation}{%
\section{Derivation of the full conditional and IWLS-proposal
distributions used for posterior
estimation}\label{derivation-of-the-full-conditional-and-iwls-proposal-distributions-used-for-posterior-estimation}}

\hypertarget{app:JMPAMM:GibbsLong}{%
\subsection{Longitudinal effects}\label{app:JMPAMM:GibbsLong}}

Let \(\boldsymbol{\gamma}_{\text{l}, k}\) be the coefficients of one of
\(k = 1, \dots, p_\text{l}\) effects in the longitudinal predictor with
a prior as given in \eqref{rappl:prior} and let further denote
\(\boldsymbol{\eta}_\text{l, -k} = \boldsymbol{\eta}_\text{l} - \boldsymbol{\eta}_\text{l, k}\),
i.e.~the longitudinal predictor without the \(k^{th}\) element. Then the
derivation of the full conditionals for this effect follows as:

\begin{align*}
p(\boldsymbol{\gamma}_{\text{l}, k} \mid \cdot) & \propto p(\boldsymbol{\gamma}_\text{l} \mid \sigma^2_\varepsilon, \sigma^2_{\boldsymbol{\gamma}_\text{l}}) \; p(\boldsymbol{y} \mid \boldsymbol{\eta}_\text{l}, \boldsymbol{\eta}_\text{ls}, \cdot)\\
& \propto \exp \left\{ -\frac{1}{2\sigma^2_{\boldsymbol{\gamma}_\text{l}}} \boldsymbol{\gamma}_{\text{l}, k}' \boldsymbol{K}_{\text{l}, k} \boldsymbol{\gamma}_{\text{l}, k}  \right\}\\
& \color{white} \propto \color{black} \exp \left\{ -\frac{1}{2\sigma^2_\varepsilon} (\boldsymbol{Z}_{\text{l}, k}\boldsymbol{\gamma}_{\text{l}, k} - (\boldsymbol{y} - \boldsymbol{\eta}_{\text{l}, -k} - \boldsymbol{\eta}_\text{ls}))'(\boldsymbol{Z}_{\text{l}, k}\boldsymbol{\gamma}_{\text{l}, k} - (\boldsymbol{y} - \boldsymbol{\eta}_{\text{l}, -k} - \boldsymbol{\eta}_\text{ls})) \right\}\\
\boldsymbol{\gamma}_{\text{l}, k} \mid \cdot & \sim \text{N}(\mu^*_{\boldsymbol{\gamma}_{\text{l}, k}}, \Sigma^*_{\boldsymbol{\gamma}_{\text{l}, k}}) \\
\Sigma^*_{\boldsymbol{\gamma}_{\text{l}, k}} & = \left(\frac{1}{\sigma^2_\varepsilon}\boldsymbol{Z}_{l, k}'\boldsymbol{Z}_{l, k} + \frac{1}{\sigma^2_{\boldsymbol{\gamma}_{\text{l}, k}}} \boldsymbol{K}_{\text{l}, k}\right)^{-1}\\
\mu^*_{\boldsymbol{\gamma}_{\text{l}, k}} &= \Sigma^*_{\boldsymbol{\gamma}_{\text{l}, k}} \left(\frac{1}{\sigma^2_\varepsilon}\left(\boldsymbol{Z}_{l, k}'(\boldsymbol{y} - \boldsymbol{\eta}_\text{l, -k} - \boldsymbol{\eta}_\text{ls})\right) +  \frac{1}{\sigma^2_{\boldsymbol{\gamma}_{\text{l}, k}}} \boldsymbol{K}_{\text{l}, k} \right)
\end{align*}

\hypertarget{app:JMPAMM:IWLSsurv}{%
\subsection{Survival effects}\label{app:JMPAMM:IWLSsurv}}

Since the full conditional distribution of the \(k^{th}\) survival
specific coefficients
\(p(\boldsymbol{\gamma}_{\text{s}, k} \mid \boldsymbol{\delta}, \sigma^2_{\boldsymbol{\gamma}_{\text{s}, k}}, \cdot)\)
out of \(k = 1, \dots, p_\text{s}\) survival specific effects are
analytically intractable, we use MH-steps with IWLS proposals, which
approximate the true log-full conditionals. Consider the (standard) full
conditional \begin{align*}
p(\boldsymbol{\gamma}_{\text{s}, k} \mid \boldsymbol{\delta}, \sigma^2_{\boldsymbol{\gamma}_{\text{s}, k}}, \cdot) & \propto p(\boldsymbol{\gamma}_{\text{s}, k} \mid \sigma^2_{\boldsymbol{\gamma}_{\text{s}, k}}) \; p(\boldsymbol{\delta} \mid \boldsymbol{\eta}_\text{s}, \boldsymbol{\eta}_\text{ls}, \cdot).
\end{align*}

Let \(\boldsymbol{Z}_{\text{s}, k}\) be the corresponding design matrix
of effect \(\boldsymbol{\gamma}_{\text{s}, k}\) and
\(\frac{1}{\sigma^2_{\boldsymbol{\gamma}_{\text{s}, k}}} \boldsymbol{K}\)
the variance statement of prior
\(p(\boldsymbol{\gamma}_{\text{s}, k} \mid \sigma^2_{\boldsymbol{\gamma}_{\text{s}, k}})\)
(compare prior given in \eqref{rappl:prior}). Then draw IWLS proposal
\(\boldsymbol{\gamma}_{\text{s}, k}^*\) from a normal distribution
density
\(q(\boldsymbol{\gamma}_{\text{s}, k}^* \mid \boldsymbol{\gamma}_{\text{s}, k}^{[t]})\)
with \(\boldsymbol{\gamma}_{\text{s}, k}^{[t]}\) being the value of
\(\boldsymbol{\gamma}_{\text{s}, k}\) at iteration \(t\) of the MCMC
algorithm. More specifically
\[\boldsymbol{\gamma}_{\text{s}, k}^* \sim N\left(\boldsymbol{\mu}^{[t]}_{\boldsymbol{\gamma}_{\text{s}, k}}, \boldsymbol{P}^{[t]-1}_{\boldsymbol{\gamma}_{\text{s}, k}}\right) 
\] \[
\text{with} \quad 
\boldsymbol{P}^{[t]}_{\boldsymbol{\gamma}_{\text{s}, k}} = \boldsymbol{Z}_{\text{s}, k}'\boldsymbol{W}^{[t]}_\text{s}\boldsymbol{Z}_{\text{s}, k} + \frac{1}{\sigma^{2[t]}_{\boldsymbol{\gamma}_{\text{s}, k}}} \boldsymbol{K}_{\boldsymbol{\gamma}_{\text{s}, k}} 
\quad \text{and} \quad 
\boldsymbol{\mu}^{[t]}_{\boldsymbol{\gamma}_{\text{s}, k}} = \left(\boldsymbol{P}^{[t]}_{\boldsymbol{\gamma}_{\text{s}, k}}\right)^{-1} \boldsymbol{Z}_{\text{s}, k}'\boldsymbol{W}^{[t]}_\text{s}\left(\tilde{\boldsymbol{y}}^{[t]}_\text{s} - \boldsymbol{\eta}^{[t]}_{\text{s},-k}\right).\]
Here
\(\boldsymbol{\eta}^{[t]}_{\text{s},-k} = \boldsymbol{\eta}^{[t]}_{\text{s}} - \boldsymbol{\eta}^{[t]}_{\text{s},k}\),
and \(\boldsymbol{W}_{\text{s}}^{[t]}\) denotes the working weights and
\(\tilde{\boldsymbol{y}}_{\text{s}}^{[t]}\) the working observations all
evaluated at the current state \(t\) of the MCMC chain. The definition
of working weights and observations is given further below.\\
The acceptance probability of the IWLS proposal
\(\boldsymbol{\gamma}^*\) is then
\[\alpha\left(\boldsymbol{\gamma}_{\text{s}, k}^* \mid \boldsymbol{\gamma}_{\text{s}, k}^{[t]}\right) = \min \left\{\frac{L\left(\boldsymbol{\gamma}_{\text{s}, k}^*\right) \, p\left(\boldsymbol{\gamma}_{\text{s}, k}^*\right) \, q\left(\boldsymbol{\gamma}_{\text{s}, k}^{[t]} \mid \boldsymbol{\gamma}_{\text{s}, k}^*\right)}{L\left(\boldsymbol{\gamma}_{\text{s}, k}^{[t]}\right) \, p\left(\boldsymbol{\gamma}_{\text{s}, k}^{[t]}\right) \, q\left(\boldsymbol{\gamma}_{\text{s}, k}^* \mid \boldsymbol{\gamma}_{\text{s}, k}^{[t]}\right)}, 1 \right\}\]
with
\(L(\boldsymbol{\gamma}_{\text{s}, k}) = p(\boldsymbol{\delta} \mid \boldsymbol{\gamma}_{\text{s}, k}, \cdot)\)
being the likelihood evaluated at the proposal
\(\boldsymbol{\gamma}^*_{\text{s}, k}\) as well as the current state
\(\boldsymbol{\gamma}^{[t]}\) of the effect.\\
If the proposal then is accepted it becomes the new state
\(\boldsymbol{\gamma}_{\text{s}, k}^{[t + 1]} = \boldsymbol{\gamma}_{\text{s}, k}^*\),
otherwise the current state remains
\(\boldsymbol{\gamma}_{\text{s}, k}^{[t + 1]} = \boldsymbol{\gamma}_{\text{s}, k}^{[t]}\).
Acceptance is established via random draws from a uniform distribution
following the logic:

\begin{enumerate}
\def\labelenumi{\arabic{enumi}.}
\tightlist
\item
  Draw \(u \sim \text{Unif}(0,1)\)\\
\item
  If \(u \leq \alpha\)\\
  then \(\boldsymbol{\gamma}^{[t + 1]} = \boldsymbol{\gamma}^*\)\\
  else \(\boldsymbol{\gamma}^{[t + 1]} = \boldsymbol{\gamma}^{[t]}\).
\end{enumerate}

For the definition of the working weights and observations consider the
log-full conditional \[
\log (p(\boldsymbol{\gamma}_{\text{s}, k} \mid \boldsymbol{\delta}, \sigma^2_{\boldsymbol{\gamma}_{\text{s}, k}}, \cdot))\propto -\frac{1}{2\sigma^2_{\boldsymbol{\gamma}_{\text{s}, k}}} \boldsymbol{\gamma}'_{\text{s}, k} \boldsymbol{K}_{\boldsymbol{\gamma}_{\text{s}, k}} \boldsymbol{\gamma}_{\text{s}, k} + \ell(\boldsymbol{\eta}_\text{s}),
\] where \(\ell(\boldsymbol{\eta}_\text{s})\) denotes the log-likelihood
depending on predictor
\(\boldsymbol{\eta}_\text{s} = \sum_{k=1}^{p_\text{s}} \boldsymbol{Z}_{\text{s}, k} \boldsymbol{\gamma}_{\text{s}, k}\)
(compare \eqref{matrixNot}) thus including
\(\boldsymbol{\gamma}_{\text{s}, k}\). Further, define the score vector
\(\boldsymbol{v}_{\text{s}}\) as \[
\boldsymbol{v}^{[t]}_{\text{s}} = \frac{\partial\ell(\boldsymbol{\eta}^{[t]}_\text{s})}{\partial \boldsymbol{\eta}^{[t]}_\text{s}},
\] i.e.~the vector of first derivatives of
\(\ell(\boldsymbol{\eta}_\text{s})\) with respect to the predictor
\(\boldsymbol{\eta}_\text{s}\) evaluated at the current iteration \(t\)
and the working weights also evaluated at iteration \(t\) as
\[\boldsymbol{W}^{[t]}_\text{s} = \text{diag} \left(w_1 \left(\eta_{\text{s},1}^{[t]} \right), \dots, w_{n_a}\left(\eta_{\text{s},n_a}^{[t]} \right)\right)\]
with \(n_a\) the number of observations in the augmented dataset
(compare Section \ref{sec:PAMM} Table \ref{tab:dataug}) and
\[w_i\left(\eta_{\text{s}, i}^{[t]}\right) = -\text{E} \left( \frac{\partial\ell(\eta^{[t]}_{\text{s}, i})}{\partial^2 \eta^{[t]}_{\text{s},i}} \right) = -\text{E} \left( \frac{\partial v^{[t]}_{\text{s}, i}}{\partial \eta^{[t]}_{\text{s},i}} \right).\]\\
The vector of working observations
\(\tilde{\boldsymbol{y}}^{[t]}_\text{s} = \left(\tilde{y}_{\text{s}1}\left((\eta_1^{[t]}\right), \dots, \tilde{y}_{\text{s}n_a}\left(\eta_{n_a}^{[t]}\right)\right)'\)
is then determined by
\[\tilde{\boldsymbol{y}}^{[t]}_\text{s} = \boldsymbol{\eta}^{[t]}_\text{s} + \left(\boldsymbol{W}_\text{s}^{[t]}\right)^{-1}\boldsymbol{v}^{[t]}_{\text{s}}.\]

\hypertarget{app:JMPAMM:IWLSshared}{%
\subsection{Shared effects}\label{app:JMPAMM:IWLSshared}}

The full conditional of the \(k^{th}\) coefficient
\(p(\boldsymbol{\gamma}_{\text{ls}, k} \mid \boldsymbol{y}, \boldsymbol{\delta}, \sigma^2_{\boldsymbol{\gamma}_{\text{ls}, k}}, \cdot)\)
of the \(k = 1, \dots, p_\text{ls}\) shared effects are neither
tractable. Therefore, we also apply an MH-step with IWLS-proposal here.
The procedure is similar to the survival specific coefficients but needs
to consider the joint likelihood of both model parts. First consider the
full conditional
\[p\left(\boldsymbol{\gamma}_\text{ls, k} \mid \cdot \right) \propto p(\boldsymbol{\gamma}_\text{ls, k} \mid \sigma^2_{\boldsymbol{\gamma}_\text{ls}}) \; p(\boldsymbol{y} \mid \boldsymbol{\eta}_\text{l}, \boldsymbol{\eta}_{\text{ls}}, \cdot) \; p(\boldsymbol{\delta} \mid \boldsymbol{\eta}_\text{s}, \boldsymbol{\eta}_{\text{ls}}, \cdot).\]

Let \(\boldsymbol{Z}_{\text{ls},k}\) be the corresponding design matrix
of effect \(\boldsymbol{\gamma}_{\text{ls}, k}\) and
\(\frac{1}{\sigma^2_{\boldsymbol{\gamma}_{\text{ls}, k}}} \boldsymbol{K}\)
the variance statement of prior
\(p(\boldsymbol{\gamma}_{\text{ls}, k} \mid \sigma^2_{\boldsymbol{\gamma}_{\text{ls}, k}})\)
(compare prior given in \eqref{rappl:prior}). Now to approximate the
full conditional we draw IWLS proposal
\(\boldsymbol{\gamma}_{\text{ls}, k}^*\) from a normal distribution
density
\(q(\boldsymbol{\gamma}_{\text{ls}, k}^* \mid \boldsymbol{\gamma}_{\text{ls}, k}^{[t]})\)
with \(\boldsymbol{\gamma}_{\text{ls}, k}^{[t]}\) being the value of
\(\boldsymbol{\gamma}_{\text{ls}, k}\) at iteration \(t\) of the MCMC
algorithm. More specifically
\[\boldsymbol{\gamma}_{\text{ls}, k}^* \sim N\left(\boldsymbol{\mu}^{[t]}_{\boldsymbol{\gamma}_{\text{ls}, k}}, \left(\boldsymbol{P}^{[t]}_{\boldsymbol{\gamma}_{\text{ls}, k}}\right)^{-1}\right)
\]
\[\text{with} \quad \boldsymbol{P}^{[t]}_{\boldsymbol{\gamma}_{\text{ls}, k}} = \boldsymbol{Z}_{\text{ls},k}'\boldsymbol{W}_\text{ls}^{[t]}\boldsymbol{Z}_{\text{ls},k} + \frac{1}{\sigma^{2[t]}_{\boldsymbol{\gamma}_{\text{ls}, k}}} \boldsymbol{K}_{\boldsymbol{\gamma}_{\text{ls},k}} \quad \text{and} \quad \boldsymbol{\mu}^{[t]}_{\boldsymbol{\gamma}_{\text{ls}, k}} = \left(\boldsymbol{P}^{[t]}_{\boldsymbol{\gamma}_{\text{ls}, k}}\right)^{-1} \boldsymbol{Z}_{\text{ls},k}'\boldsymbol{W}_\text{ls}^{[t]}\left(\tilde{y}_\text{ls}^{[t]} - \boldsymbol{\eta}_{\text{ls},-k} ^{[t]} \right).\]
The rest of the algorithm is analogous to the survival effects.

To see how the working weights and observations build for the
coefficients in the shared predictor, consider first the log-full
conditional \[
\log (p(\boldsymbol{\gamma}_{\text{ls}, k} \mid \boldsymbol{y}, \boldsymbol{\delta}, \sigma^2_{\boldsymbol{\gamma}_{\text{ls}, k}}, \cdot))\propto -\frac{1}{2\sigma^2_{\boldsymbol{\gamma}_{\text{ls}, k}}} \boldsymbol{\gamma}'_{\text{ls}, k} \boldsymbol{K}_{\boldsymbol{\gamma}_{\text{ls},k}} \boldsymbol{\gamma}_{\text{ls}, k} + \ell_y(\boldsymbol{\eta}_\text{ls}) + \ell_\delta(\boldsymbol{\eta}_\text{ls}),
\] where \(\ell_y(\boldsymbol{\eta}_\text{ls})\) denotes the
longitudinal part of the log-likelihood and
\(\ell_\delta(\boldsymbol{\eta}_\text{ls})\) the survival/ poisson part
of the log-likelihood depending on predictor
\(\boldsymbol{\eta}_\text{ls} = \sum_{k=1}^{p_\text{ls}} \boldsymbol{Z}_{\text{ls}, k} \boldsymbol{\gamma}_{\text{ls}, k}\)
(compare \eqref{matrixNot}) thus including
\(\boldsymbol{\gamma}_{\text{ls}, k}\).

The vector of scores, i.e.~first derivatives of the log-likelihoods with
respect to \(\boldsymbol{\eta}_\text{ls}\) evaluated at iteration \(t\),
is \[
\boldsymbol{v}^{[t]}_{\text{ls}} = \boldsymbol{v}^{[t]}_{y,\text{ls}} + \alpha\boldsymbol{v}^{[t]}_{\delta,\text{ls}} = \frac{\partial\ell_y(\boldsymbol{\eta}^{[t]}_\text{ls})}{\partial \boldsymbol{\eta}^{[t]}_\text{ls}} + \frac{\partial\ell_\delta(\boldsymbol{\eta}^{[t]}_\text{ls})}{\partial \boldsymbol{\eta}^{[t]}_\text{ls}}.
\] The working weights evaluated at iteration \(t\) can then be derived
as
\[\boldsymbol{W}^{[t]}_\text{ls} = \text{diag}\left(w_1\left(\eta_{\text{ls},1}^{[t]}\right), \dots, w_{n_a}\left(\eta_{\text{ls},n_a}^{[t]}\right)\right)\]
with \(n_a\) the number of observations in the augmented dataset and
\begin{align*}
w_i(\eta_{\text{ls}, i}^{[t]}) &= -\text{E} \left( \frac{\partial v^{[t]}_{\text{ls}, i}}{\partial \eta^{[t]}_{\text{ls},i}} \right) = -\text{E} \left( \frac{\partial\boldsymbol{v}^{[t]}_{y,\text{ls}, i} + \alpha\boldsymbol{v}^{[t]}_{\delta,\text{ls}, i}}{\partial \eta^{[t]}_{\text{ls},i}} \right)= 
- \text{E} \left(\frac{\partial\ell_y(\boldsymbol{\eta}^{[t]}_{\text{ls},i})}{\partial^2 \boldsymbol{\eta}^{[t]}_{\text{ls},i}} \right) - \text{E} \left(\frac{\partial\ell_\delta(\boldsymbol{\eta}^{[t]}_{\text{ls},i})}{\partial^2 \boldsymbol{\eta}^{[t]}_{\text{ls},i}} \right) \\
&= w_{y,i}(\eta_{\text{ls}, i}^{[t]}) + \alpha^2 w_{\delta, i}(\eta_{\text{ls}, i}^{[t]}).
\end{align*}

The working observations then follow analogously as \[
\tilde{\boldsymbol{y}}_\text{ls}^{[t]} = \boldsymbol{\eta}^{[t]}_\text{ls} + \left(\boldsymbol{W}_\text{ls}^{[t]}\right)^{-1}\boldsymbol{v}^{[t]}_{\text{ls}}.
\]

\hypertarget{variances}{%
\subsection{Variances}\label{variances}}

\hypertarget{model-variance}{%
\subsubsection{Model variance}\label{model-variance}}

Let \(N = \sum_{i = 1}^{n} n_i\) be the total number of longitudinal
observations as the sum of all observations \(n_i\) per individual \(i\)
across all individuals \(n\). Then the full conditional of the model
variance follows as \begin{align*}
p(\sigma^2_\varepsilon \mid \cdot) & \propto p(\sigma^2_\varepsilon) \; p(\boldsymbol{y} \mid \boldsymbol{\eta}_\text{l}, \boldsymbol{\eta}_\text{ls}, \sigma^2_\varepsilon) \\
& \propto (\sigma^2_\varepsilon)^{-a_0 -1} \; \exp\left\{-\textstyle{\frac{b_0}{\sigma^2_\varepsilon}}\right\} \\
& \color{white} \propto \color{black} \left(\sigma^2_\varepsilon\right)^{-\frac{N}{2}} \; \exp\left\{-\textstyle{\frac{1}{2\sigma^2_\varepsilon}} (\boldsymbol{y} - \boldsymbol{\eta}_\text{l} - \boldsymbol{\eta}_\text{ls})'(\boldsymbol{y} - \boldsymbol{\eta}_\text{l} - \boldsymbol{\eta}_\text{ls})\right\} \\
& \propto \left(\sigma^2_\varepsilon\right)^{-(a_0 + \frac{N}{2}) - 1} \; \exp\left\{-\textstyle{\frac{1}{\sigma^2_\varepsilon}} (b_0 + \textstyle{\frac{1}{2}} (\boldsymbol{y} - \boldsymbol{\eta}_\text{l} - \boldsymbol{\eta}_\text{ls})'(\boldsymbol{y} - \boldsymbol{\eta}_\text{l} - \boldsymbol{\eta}_\text{ls}))\right\}\\
\sigma^2_\varepsilon \mid \cdot & \sim \text{IG}\left(a_0 + \textstyle{\frac{N}{2}},  b_0 + (\boldsymbol{y} - \boldsymbol{\eta}_\text{l} - \boldsymbol{\eta}_\text{ls})'(\boldsymbol{y} - \boldsymbol{\eta}_\text{l} - \boldsymbol{\eta}_\text{ls})\right)
\end{align*}

\hypertarget{variance-of-coefficients}{%
\subsubsection{Variance of
coefficients}\label{variance-of-coefficients}}

With \(p_{\cdot}\) being the number of covariates in each
predictor(longitudinal, shared, survival) the full conditional of the
\(k_{\cdot} = 1, \dots, p_{\cdot}\) variances of the corresponding
effects is \begin{align*}
p(\sigma^2_{\boldsymbol{\gamma}_{k_\cdot}} \mid \cdot) & \propto p(\sigma^2_{\boldsymbol{\gamma}_{k_\cdot}}) \; p(\boldsymbol{\gamma}_{k_\cdot} \mid \sigma^2_{\boldsymbol{\gamma}_{k_\cdot}}) \\ 
& \propto (\sigma^2_{\boldsymbol{\gamma}_{k_\cdot}})^{-a_{k_\cdot} - 1} \; \exp\left\{-\textstyle{\frac{b_{k_\cdot}}{\sigma^2_{\boldsymbol{\gamma}_{k_\cdot}}}}\right\} \\
& \color{white} \propto \color{black} \left(\sigma^2_{\boldsymbol{\gamma}_{k_\cdot}}\right)^{-\text{rk}\boldsymbol{K}_{k_\cdot}} \; \exp\left\{-\textstyle{\frac{1}{2\sigma^2_{\boldsymbol{\gamma}_{k_\cdot}}}} \boldsymbol{\gamma}_{\cdot, k_{\cdot}}' \boldsymbol{K}_{k_{\cdot}} \boldsymbol{\gamma}_{\cdot, k_{\cdot}} \right\} \\
& \propto \left(\sigma^2_{\boldsymbol{\gamma}_{k_\cdot}}\right)^{-(a_{k_\cdot} + \text{rk}\boldsymbol{K}_{k_\cdot}) - 1} \; \exp\left\{-\textstyle{\frac{1}{\sigma^2_{\boldsymbol{\gamma}_{k_\cdot}}}} (b_{k_\cdot} + \boldsymbol{\gamma}_{\cdot, k_{\cdot}}' \boldsymbol{K}_{k_{\cdot}} \boldsymbol{\gamma}_{\cdot, k_{\cdot}})\right\} \\
\sigma^2_{\boldsymbol{\gamma}_{k_\cdot}} \mid \cdot & \sim \text{IG}\left(a_{k_\cdot} + \text{rk}\boldsymbol{K}_{k_\cdot},  b_{k_\cdot} + \boldsymbol{\gamma}_{\cdot, k_{\cdot}}' \boldsymbol{K}_{k_{\cdot}} \boldsymbol{\gamma}_{\cdot, k_{\cdot}} \right)
\end{align*}

\newpage

\hypertarget{detailed-results-of-simulations-study}{%
\section{Detailed results of simulations
study}\label{detailed-results-of-simulations-study}}

Figure \ref{fig:appres} displays the results of the simulation study
detailed by individual effect.

\begin{figure}[h!]

{\centering \includegraphics[height=13.4 cm,angle= 90]{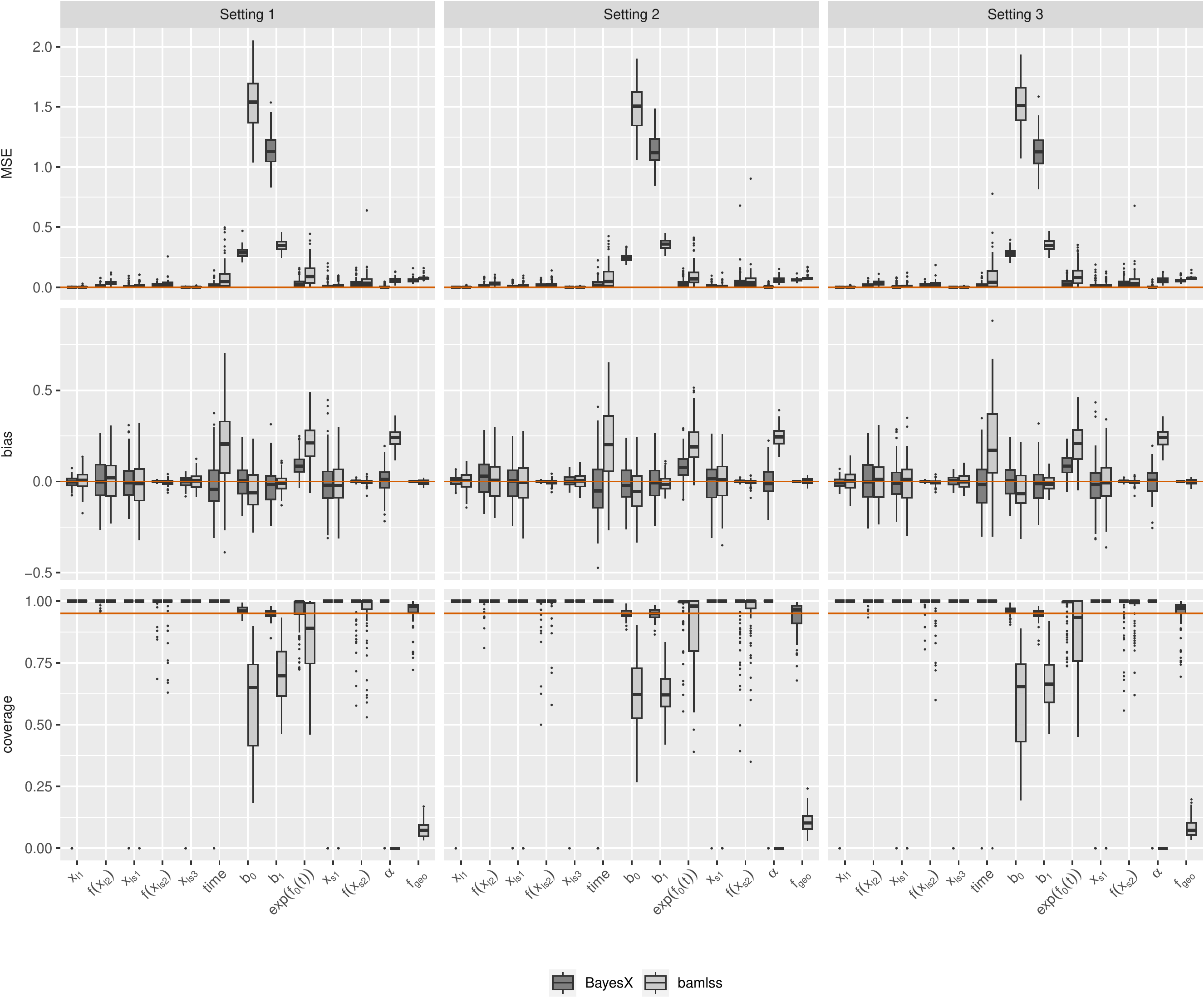} 

}

\caption{Boxplots of mean squared error (MSE), bias and 95$\%$-coverage per method by effect and simulation setting (Setting 1 - $f_{geo}$ in $\eta_{ls}$, Setting 2 - $f_{geo}$ in $\eta_{s}$, Setting 3 - $f_{geo}$ in $\eta_{l}$). The orange horizontal line marks the reference value of each statistic.\label{fig:appres}}\label{fig:appres}
\end{figure}

\newpage

\hypertarget{statistical-overview-of-run-times}{%
\section{Statistical overview of run
times}\label{statistical-overview-of-run-times}}

Table \ref{tab:app_runtimes} details descriptive measures of the run
times.

\begin{table}[!h]

\caption{\label{tab:unnamed-chunk-26}Statistical overview of run times of 100 replications per setting and estimation method. \label{tab:app_runtimes}}
\centering
\begin{tabular}[t]{ccccccc}
\toprule
Method & Min. & 1st.Qu. & Median & Mean & 3rd.Qu. & Max.\\
\midrule
\addlinespace[0.3em]
\multicolumn{7}{l}{\textbf{Setting 1}}\\
\hspace{1em}BayesX & 1.30 & 1.53 & 1.58 & 1.58 & 1.62 & 1.80\\
\hspace{1em}bamlss & 3.17 & 3.30 & 3.38 & 3.86 & 3.42 & 10.62\\
\addlinespace[0.3em]
\multicolumn{7}{l}{\textbf{Setting 2}}\\
\hspace{1em}BayesX & 1.10 & 1.22 & 1.26 & 1.26 & 1.30 & 1.41\\
\hspace{1em}bamlss & 1.68 & 2.04 & 2.17 & 2.50 & 2.26 & 9.62\\
\addlinespace[0.3em]
\multicolumn{7}{l}{\textbf{Setting 3}}\\
\hspace{1em}BayesX & 1.03 & 1.25 & 1.30 & 1.28 & 1.32 & 1.40\\
\hspace{1em}bamlss & 2.01 & 2.06 & 2.21 & 2.58 & 2.26 & 12.98\\
\bottomrule
\end{tabular}
\end{table}

\bibliographystyle{unsrt}
\bibliography{references.bib}

\end{document}